\documentclass[pre,aps,twocolumn]{revtex4}
\usepackage[pdftex]{graphicx}
\usepackage{subfigure,amsbsy,amsmath}

\usepackage[latin1]{inputenc}

\vfil

\usepackage{amssymb}
\usepackage{amsfonts}
\usepackage{mathrsfs}

\begin{document}

\title{Connecting Curves in Higher Dimensions}

\author{Greg Byrne$^1$, Juan Cebral$^1$ and Robert Gilmore$^2$}

\affiliation{$^1$Center for Computational Fluid Dynamics, George Mason University, Fairfax, VA 22030, USA}
\affiliation{$^2$Physics Department, Drexel University, Philadelphia,
  Pennsylvania 19104, USA} 

\date{\today, {\it Physical Review E}: To be submitted.}

\begin{abstract}
Connecting curves have been shown to organize the
rotational structure of strange attractors in three-dimensional 
dynamical systems.  We extend the description of connecting
curves and their properties to higher dimensions within
the special class of differential dynamical systems.
The general properties of connecting curves are
derived and selection rules stated.  Examples are
presented to illustrate these properties for
dynamical systems of dimension $n=3,4,5$.

\end{abstract}

\pacs{PACS numbers: XXXXXXX}
\maketitle

\section{Introduction}

Autonomous dynamical systems are defined by equations
of the form $\dot{x}_i =f_i(x_1, x_2, \dots , x_n)$,
$i=1,2, \dots, n$.  When a bounded attractor exists,
its general structure is partly described by the location
of the fixed points $\dot{x}_i=0$. Typically the fixed points 
are isolated.  More information about the structure of an 
attracting set can be determined from the stability of each of
the fixed points.  The stability is determined from the
eigenvalues of the jacobian $J_{ij} = \partial f_i/\partial x_j$
at each fixed point, and the eigenvectors associated
with each eigenvalue.

When a dynamical system $\dot{x}_i =f_i$ generates a strange attractor, 
certain parts of the attractor ``swirl'' around one-dimensional 
invariant sets called connecting curves\cite{Gilmore10}.  In such cases, 
connecting curves organize the global structure of the flow, and therefore 
provide more information than just the location and stability of 
the fixed points.  Connecting curves defined by the eigenvalue-like 
condition $J_{ij}f_j=\lambda f_i$ have been studied in a number of 
three-dimensional dynamical systems.  They are recognized as a kind of 
skeleton that helps to define the structure of a strange attractor.

In this work we explore the properties of connecting curves
in dynamical systems of dimension $n \ge 3$.  This is done
in a restricted class of dynamical systems called
differential dynamical systems.  We describe the basic 
connectivity between fixed points and outline conditions 
that determine whether or not they lie on the connecting 
curve.  We also describe the systematics of connecting curve 
attachment to or detachment from a fixed point and the 
recombinations or reconnections that can take place 
between different connecting curves. 

Fixed points and their eigenvalue spectrum are used to predict 
changes in the local stability of a connecting curve as the 
control parameters are varied.  Under certain conditions, we 
show that the global stability of a connecting curve can also 
be predicted.  When the stability is such that the flow undergoes 
a swirling motion around the connecting curve, an integer 
index $\kappa$ is used to describe the swirling in terms of 
vortex or hypervortex structures.  The formation of these structures 
plays an important role in determining the topology of strange attractors 
in higher dimensions.

In Sec. \ref{sec:connectingcurves} we review the definition
of connecting curves and their properties.  The particular class
of dynamical systems studied is introduced in 
Sec. \ref{sec:diffdynsys}.  In that Section we also describe
how the distribution of fixed points in such systems is
systematically organized by cuspoid catastrophes $A_m'$, where
$m$ is the maximum number of fixed points allowed by variation
of the control parameters.  We also describe how the
stability properties of each fixed point are determined by
another cuspoid catastrophe $A_n'$, where $n$ is the
dimension of the dynamical system.
In Secs. \ref{sec:3D}, \ref{sec:4D}, and \ref{sec:5D} 
we study cases of three-, four-, and five-dimensional
dynamical systems with one, two, and three fixed points.  
These illustrate many of the properties
of connecting curves described in Secs. \ref{sec:connectingcurves}
and \ref{sec:diffdynsys}.  Results are summarized in
Sec. \ref{sec:conclusions}.

\section{Connecting Curves}
\label{sec:connectingcurves}

Special points exist in the phase space $R^n$ of autonomous dynamical systems 
$\dot{x}_i = f_i(x)$ where the acceleration 
${\bf \dot{f}}$ is proportional to the velocity ${\bf f}$.  These points 
satisfy the eigenvalue equation 
${\bf \dot{f}}=J{\bf f} = \lambda {\bf f}$, where 
$J = \partial f_i/\partial x_j$ is the jacobian.  
For an $n$-dimensional dynamical system, the equations represented 
by $J{\bf f} = \lambda {\bf f}$ provide $n$ constraints in an 
$n+1$ dimensional space $R^{n+1}$ consisting of $n$ phase space coordinates 
and one ``eigenvalue'': $(x_1, \dots, x_n;\lambda)$.  The intersections of the manifolds defined by these equations are one-dimensional sets
in $R^{n+1}$ whose projection into the coordinate 
subspace $R^n$ is called a connecting curve.

Much like fixed points, connecting curves provide constraints on the 
behavior of local phase space trajectories.  This behavior is determined
by examining the eigenvalue spectrum of the jacobian matrix along the length 
of the connecting curve.  The eigenvalues can occur in a variety of ways 
depending on the dimension, $n$, of the phase 
space: $n-2\kappa$ real and $\kappa$ complex conjugate 
pairs, with $\kappa=0,1, \cdots, [n/2]$.  

Subsets of a connecting curve along which $\kappa=0$ are 
called strain curves because small volumes in their vicinity 
undergo irrotational deformation under the action of the flow.  
Deformation takes place along the principle stable and unstable 
directions indicated by the eigenvectors $\xi_{R}$ of the 
real eigenvalues $\lambda_{R}$.  

In three dimensions $(n=3)$, subsets of the 
connecting curve along which $\kappa=1$ are known as 
vortex core curves.  Vortex core curves were originally 
developed to identify vortices in complex 
hydrodynamic flows\cite{Roth98,Peikert00}.  They 
were later shown to organize the large-scale 
structure of strange attractors produced 
by three-dimensional dynamical 
systems\cite{Gilmore10}.

A vortex can be decomposed into its rotational and 
non-rotational components by linearizing the flow around 
its core curve.  Rotation takes place on a plane spanned by the 
eigenvectors $\xi_{C}$ of the complex eigenvalues $\lambda_{C}$.  
The swirling flow is then transported along the eigenvector 
$\xi_{R}$ associated with the real eigenvalue $\lambda_{R}$.  
Under this combined action, trajectories are expected to undergo a 
tornado-like motion that spirals around the core in the direction 
of the flow.  

The concept of a vortex can also be extended to higher 
dimensions in the phase space of dynamical systems.  
The basic idea is the same as in three-dimensions.  
Linearized flow around the core line is resolved 
into $\kappa \geq 1$ orthogonal
planes of rotation that can be transported 
along $n-2\kappa$ directions.  We refer to higher 
dimensional vortices with $\kappa \ge 2$ as hypervortices.  
Hypervortices and their associated core curves play an important 
role in organizing the large-scale structure of strange attractors in 
higher dimensions.

At a fixed point the initial conditions for a core
curve are the eigenvectors with real eigenvalues.
This observation has a number of important consequences.
At a fixed point with $n-2\kappa$ real eigenvalues,
$n-2\kappa$ connecting curves pass through the 
fixed point.  This means that if $n-2\kappa=0$ at a 
fixed point, no connecting curves attach to the
fixed point.  If, under control parameter variation,
the stability properties of a fixed point change when a 
pair of real eigenvalues become degenerate and transform to a complex
conjugate pair, then two connecting curves must disconnect from
the fixed point.

For the remainder of this work, we assign each of the connecting 
curve subsets a unique color.  Strain curves ($\kappa=0$) are plotted in red, 
vortex core curves ($\kappa=1$) are plotted in black and 
hypervortex core curves ($\kappa=2$) are plotted in blue.

\section{Differential Dynamical Systems}
\label{sec:diffdynsys}

We use differential dynamical systems to study connecting curves in 
higher dimensions for several reasons: (a) their canonical form has 
the same structure in all dimensions,  (b) the jacobian matrix is simple 
and can be put into a Jordan-Arnol'd canonical form 
\cite{Arnold71,Gilmore81} when evaluated at the 
fixed points, and (c) only a single forcing function,  
$F(x_1,x_2, \cdots , x_{n-1},x_n;c)$, needs to be modeled.  Each of these 
properties is described in detail below.

\subsection{The Canonical Form}

Differential dynamical systems assume a canonical form in which each 
phase space coordinate except the first is the time derivative of the 
previous: $x_{i+1}=\dot{x}_i$

\begin{equation}  \begin{array}{lcl}
\dot{x}_1 &=& x_2 \\
\dot{x}_2 &=& x_3 \\
 & \vdots & \\
\dot{x}_{n-1} &=& x_n \\
\dot{x}_n &=& F(x_1,x_2, \cdots , x_{n-1},x_n;c) \end{array}
\label{eq:differentialform}
\end{equation}
This form is encountered when analyzing experimental data embedded using 
a differential embedding.  Such embeddings are equivalent to Takens 
time-delay embeddings using a minimum time delay.  

\subsection{Fixed Points and Stability}
\label{sec:ddsB}

The fixed points all lie along the $x_1$ axis because 
$\dot{x}_i=0 \Rightarrow x_{i+1}=0$ for $i=1,2, \dots ,n-1$. 
The stability properties of a fixed point at
${\bf x}_{f}=(x_{f},0 \cdots , 0,0)$
are determined by the eigenvalues of the jacobian matrix

\begin{equation}
J = 
\left[  \begin{array}{cccccc}
0 & 1 & 0 & 0 & & 0 \\
0 & 0 & 1 & 0 & & 0 \\
 & & & &\ddots & \\
0 & 0 & 0 & 0 & & 1 \\
F_1 & F_2 & F_3 & F_4 & & F_n \end{array} \right]_{{\bf x}_{f}}
\label{eq:jacobian}
\end{equation}
evaluated at the fixed point ${\bf x}_{f}$, where
as usual $F_k = \partial F/\partial x_k$.
The secular equation for the eigenvalues at the
fixed point is

\begin{equation}
\lambda^n-F_n\lambda^{n-1} \cdots -F_2\lambda^1 -F_1\lambda^0  =
\lambda^n - \sum_{k=1}^{n} F_{k} \lambda^{k-1} = 0 
\label{eq:sec} 
\end{equation}
Setting $F_n=0$ converts the jacobian (\ref{eq:jacobian}) into the 
Jordan-Arnol'd canonical form.  In this form, the jacobian provides a 
standard unfolding for the cuspoid catastrophes $A_n$ which can be 
exploited to provide information about the stability properties of 
fixed points as the control parameters are varied
\cite{Gilmore81,Thom76, Zeeman76, Poston78}.

\subsection{Source Function}

The canonical form (\ref{eq:differentialform}) can be further simplified 
by making two assumptions about the nature of the single forcing function.  
We write the forcing function as the sum of two functions $F=G_1+G_2$

\begin{equation}
F(x_1,\cdots,x_n;c) = G_1(x_1;c_1) + G_2(x_2, \cdots ,x_n;c_2) 
\end{equation}
and split the control parameters $c$ for the original source function $F({\bf x};c)$ into two subsets, $c_1$ and $c_2$.  The functions $G_1$ and $G_2$ satisfy the following properties.  

$G_1$ is a function of only one variable, $x_1$, and control 
parameter set $c_1$.  The fixed points are given by $G_1(x_1;c_1)=0$.  
Their number and location is controlled 
by the polynomial $G_1$ and the control parameters $c_1$. 
The term $F_1=G_1'$ in the jacobian in Eq. (\ref{eq:jacobian}).  We 
choose $G_1$ to be a polynomial of degree $m$ with a maximum of 
$m$ real fixed points.  In this work we take $m=1,2,3$ with a 
single control parameter $c_1=R$.  The functions $G_1$ used in 
this work are outlined below.

For a maximum of one fixed point $(m=1)$ , we choose $G_1(x_1;R)=Rx_1$.  
The fixed point is located at the 
origin and has slope $G^{'}_{1}(x_{f})=R$.  

For a maximum of two fixed points $(m=2)$, we chose 
$G_1(x_1;R)=x_{1}^{2}-R$.  There are no real fixed points 
for $R<0$, a doubly-degenerate fixed point for $R=0$ 
and two fixed points for $R>0$.  The two real 
fixed points are created in a saddle-node bifurcation.  They 
are symmetric and located at $x_{f_1}=-\sqrt{R}$ and $x_{f_2}=\sqrt{R}$ with 
slopes $G^{'}_{1}(x_{f})=2x_f$, or $-2\sqrt{R}$ 
at $x_{f_1}$ and $+2\sqrt{R}$ at $x_{f_2}$.  

For a maximum of three fixed points $(m=3)$, we choose 
$G_1(x_1;R) =x_1^3-Rx_1$.  There is one real fixed point 
for $R\leq0$ and three for $R>0$.  The three fixed points 
are created in a pitchfork bifurcation.  One fixed point 
is located at the origin.  The other two are at $x = \pm \sqrt{R}$.  The 
critical points and their slopes are 
$(x_f,G_1'(x_f) = (-\sqrt{R},2R); (0,-R); (+\sqrt{R},2R)$.

Along the $x_1$ axis the sign of $F_1=G_1'$ alternates as
successive fixed points are encountered.  For fixed points 
with focal stability, this means an alternation of 
(stable focus - unstable outset) with (unstable focus - stable inset) 
along the $x$ axis.  If $G_1' > 0$ at a fixed point then one eigenvalue 
is positive and the other two are negative, or have
negative real part.  If $G_1' < 0$ the reverse is true:
one eigenvalue is negative and the other two are either
positive or have positive real part.  The net result is that 
if there are more than two fixed points,
for an interior fixed point that is a stable focus with 
unstable outsets, the outsets can flow to the neighboring
fixed points on its left and right, which are unstable
foci with stable insets.

%%%%%%%%%%%%%%%%%%%%%%%%%%%%%%%%%%%%%%%%%%%%%%%%%
\begin{figure}[htbp]   %%%   Fig. 1
  \centering
  \includegraphics[height=6cm,width=\columnwidth]{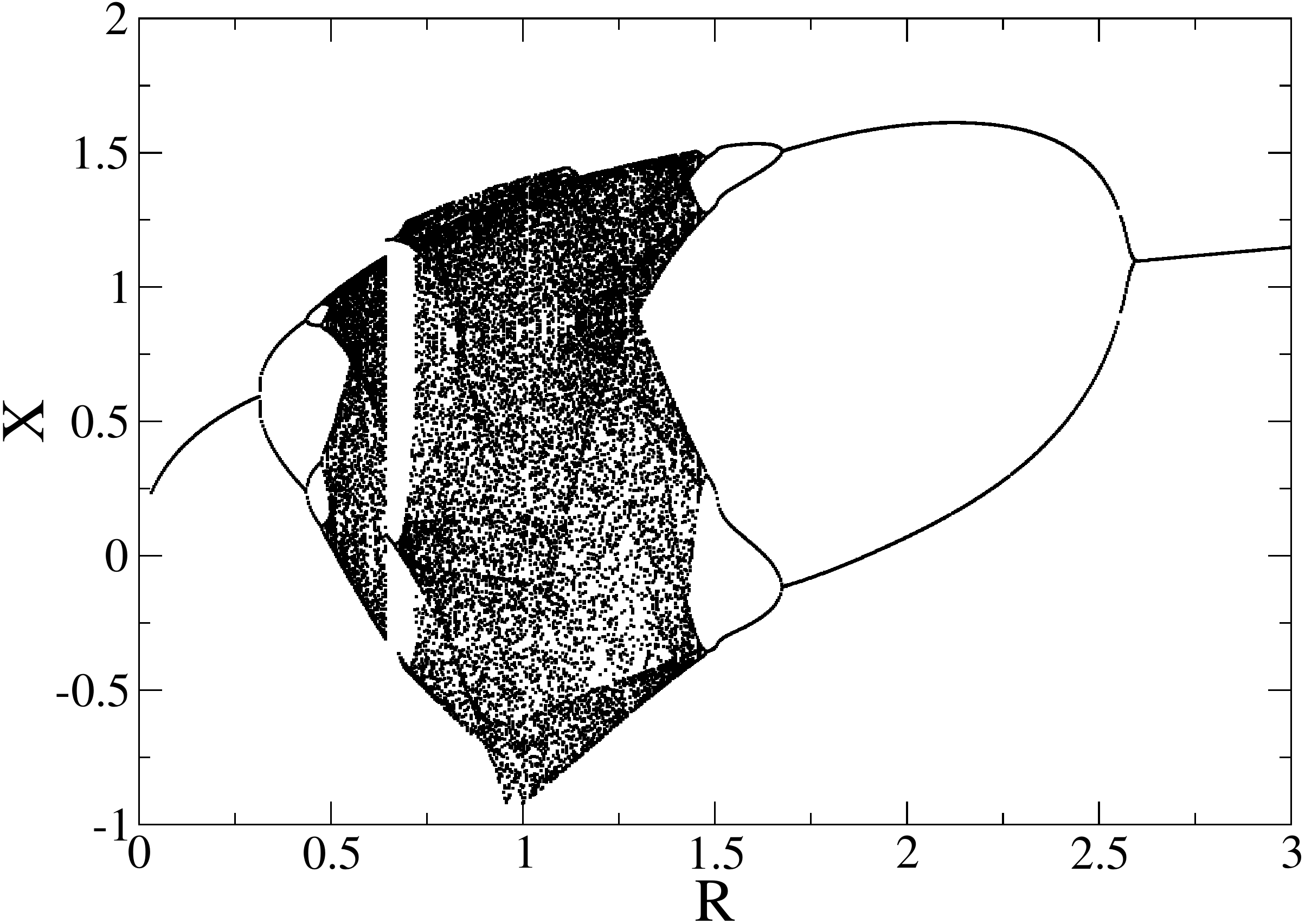}
  \caption{Bifurcation diagram created using a single Poincar\'e surface of section along the X-axis (Y=0) for Eq. (\ref{eq:3dode}) with $G_1(x,R)=x^2-R$.  A period doubling route to chaos is observed after the saddle-node bifurcation at $R=0$.  Parameter values:  $(A,B)=(-1.7,-0.8)$.}
  \label{fig:3dbif_R}    
\end{figure}
%%%%%%%%%%%%%%%%%%%%%%%%%%%%%%%%%%%%%%%%%%%%%%%%%

$G_2$ is a function of the remaining variables, 
$x_2, \cdots, x_{n-1}, x_n$.  Since the volume growth rate 
under the flow is determined by the divergence of the 
vector field, and the divergence
is $\partial F/\partial x_n = \partial G_2/\partial x_n$,
we impose the condition that $\partial G_2/\partial x_n$
is negative semi-definite throughout the phase space.  Under such an
assumption Eq. (\ref{eq:differentialform}) describes a
dissipative dynamical system.

The constant term in the Taylor series expansion of $G_2$ is zero. 
The linear terms $\sum_{j=2}^n a_jx_j$ define the terms
$F_j$ that appear in the jacobian Eq. (\ref{eq:jacobian})
when evaluated along the $x_1$ axis, specifically at
fixed points: $F_j=a_j$, $j=2, \dots, n$.  Thus, the
stability of the fixed points is determined by the
roots of the characteristic equation ${\rm det} (J-\lambda I_n)=0$,
given in Eq. (\ref{eq:sec}).  The center of gravity of the
eigenvalues at a fixed point is $F_n=(G_2)_n = a_n$.  It is
convenient to choose $a_n=0$ for two reasons.  In this case
all fixed points are unstable.  Further, Eq. (\ref{eq:sec})
assumes the form of the catastrophe function 
$G_2(x_2, \dots, x_n;s)|_{{\bf x}_{f_j}}= A_n'(\lambda;a_2, \dots, a_{n-1})$.

\section{Three Dimensions}
\label{sec:3D}

In this section we set $G_2(y,z;A,B)=Ay+Bzy^{2}$ in order to 
study three-dimensional differential dynamical systems that take the 
form 

\begin{equation}
  \begin{array}{l}
    \dot{x} = y \\
    \dot{y} = z \\
    \dot{z} = G_1(x;R)+Ay+Bzy^{2}
  \end{array}
\label{eq:3dode}
\end{equation}

The jacobian at any fixed point ${\bf x}_f$ is

\begin{equation}
J = \left[  \begin{array}{ccc}
0 & 1 & 0 \\
0 & 0 & 1 \\
G_1' & A & 0 \end{array} \right]_{{\bf x}_{f}} 
\label{eq:jcn3}
\end{equation}
with a characteristic polynomial 

\begin{equation}
{\rm det}(J-\lambda I_3))=A_3(\lambda)=\lambda^3 - A \lambda - G_1'(x_{f};R))
\label{eq:cpn3}
\end{equation}
whose roots determine the stability 
properties (focus or saddle) of the fixed 
point.  By making the substitution 
$(a,b)=(-A,-G_1'(x_{f};R))$, 
(\ref{eq:cpn3}) becomes the canonical 
unfolding of the cusp catastrophe $A_3$ 
whose bifurcation set 
is shown in Fig. \ref{fig:A3}(a).  When 
projected down into the $(a,b)$ plane, the 
bifurcation set forms a well known cusp 
shape that divides the control 
parameter space into two regions.
Inside the cusp, $(a/3)^3+(b/2)^2<0$ 
and the three eigenvalues of (\ref{eq:jcn3}) 
are real.  For control parameters in this region 
the fixed points of (\ref{eq:3dode}) have the 
stability of a saddle.  Outside the cusp, $(a/3)^3+(b/2)^2>0$ and 
and the eigenvalues of (\ref{eq:jcn3}) consist 
of one complex conjugate pair and one real.  
In this region, the fixed points have 
focal stability. 

If $G_1'=0$, the fixed point stability is determined 
along symmetry axis $b=0$.  Outside the cusp ($a>0$), the 
fixed point is a center that has two complex conjugate 
eigenvalues with zero real part and a real eigenvalue that 
takes on a value of zero.  Inside the cusp ($a<0$), the 
fixed point is a saddle and has three real eigenvalues.  Two 
of the eigenvalues differ only by a sign.  The third is equal to 
zero.  

Connecting curves satisfy the 
following two constraints in the three-dimensional 
space $(x,y,\lambda)$ (c.f., Appendix 1)

\begin{equation}
\begin{array}{rcl}
f(x,y,\lambda) &=&\lambda^2y \\
\left(G_1'(x)+(B-2A\lambda y^2)\lambda\right)y+ \\
By^2f(x,y,\lambda) &=& \lambda f(x,y,\lambda)
\end{array}
\label{eq:collapsedn3}
\end{equation}
where $f(x,y,\lambda) = G_1(x)+Ay+B\lambda y^3$.

We use polynomials $G_1(x;R)$ of degree $m=2,3$ in the sections below 
to explore the spectrum of changes that can occur in the fixed points 
and connecting curves as the parameters $R$ and $B$ are varied.

\subsection{$m=2$}
 
We begin by fixing $(A,B)=(-1.7,-0.8)$ and varying 
$R$ for $G_1(x_1;R)=x^2-R$.  A standard 
period-doubling route to chaos is observed in the 
bifurcation diagram shown in Fig. \ref{fig:3dbif_R}.  

\begin{figure}[htbp] %%%   Fig. 2
\begin{minipage}{\columnwidth}
\centering
\subfigure[Bifurcation set for the cusp catastrophe $A_3$.  The number of real eigenvalues 
in each region are indicated.]{
  \includegraphics[height=5cm,width=\columnwidth]{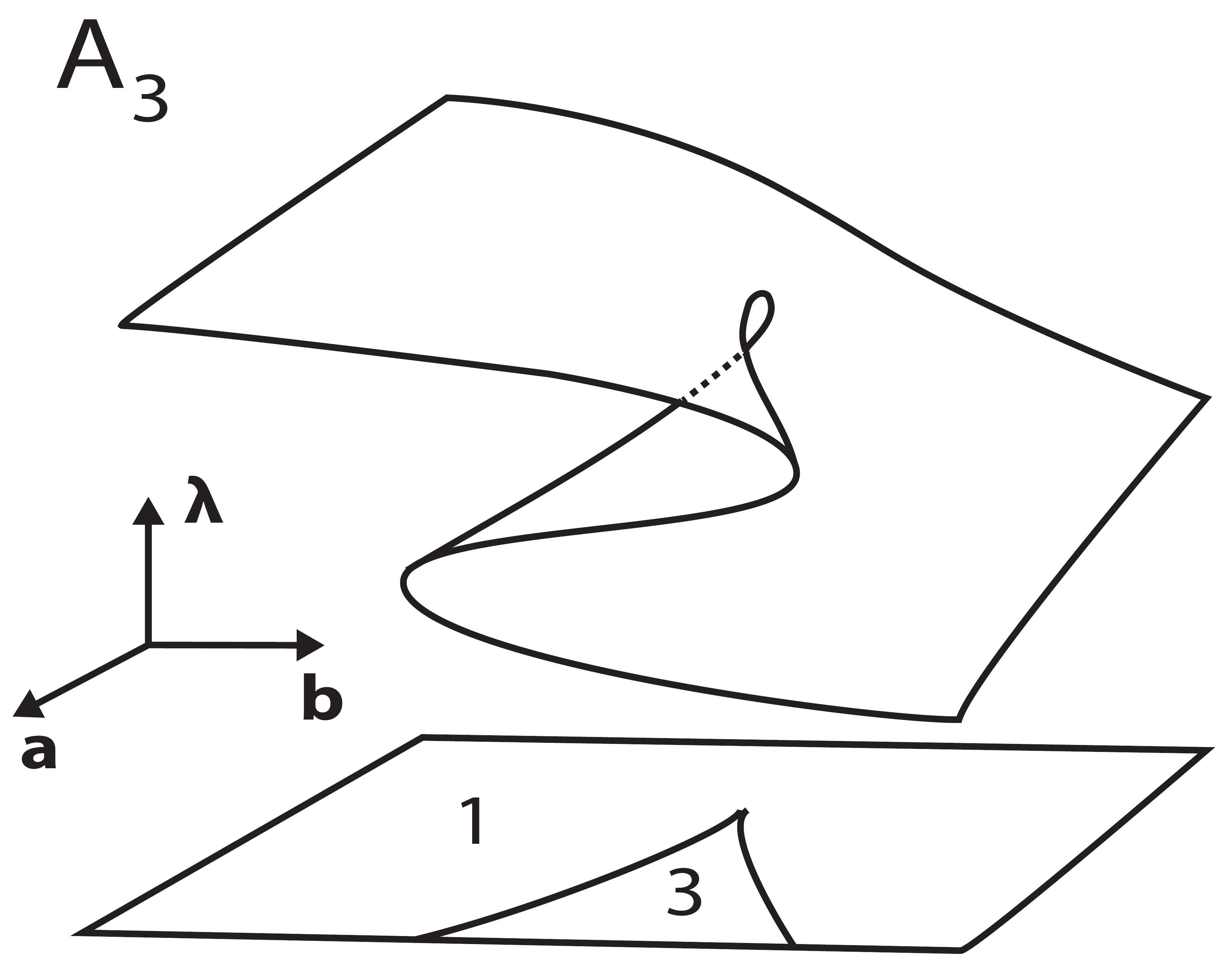}}
\subfigure[Fixed point evolution for control parameters
  $(A,B)=(-1.7,-0.8)$ for $R<0$ and $R=1$ (center and outer
  pair)]{\includegraphics[height=6cm,width=\columnwidth]{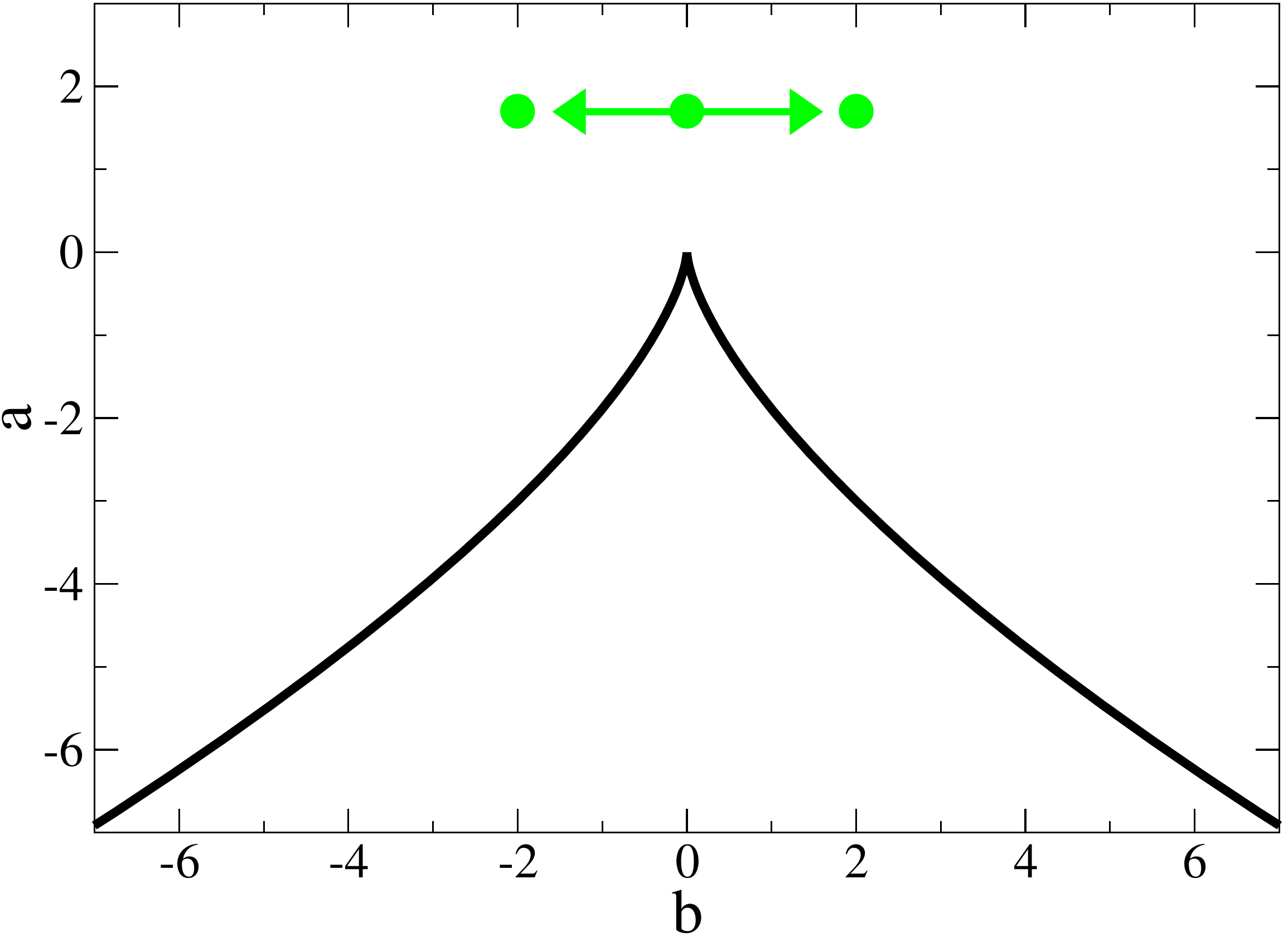}}
\subfigure[Fixed point evolution for control parameters $(B,R)=(-0.8,1)$ and $A=(1.0617,2.642,5.9012)$ (top to bottom)]{\includegraphics[height=6cm,width=\columnwidth]{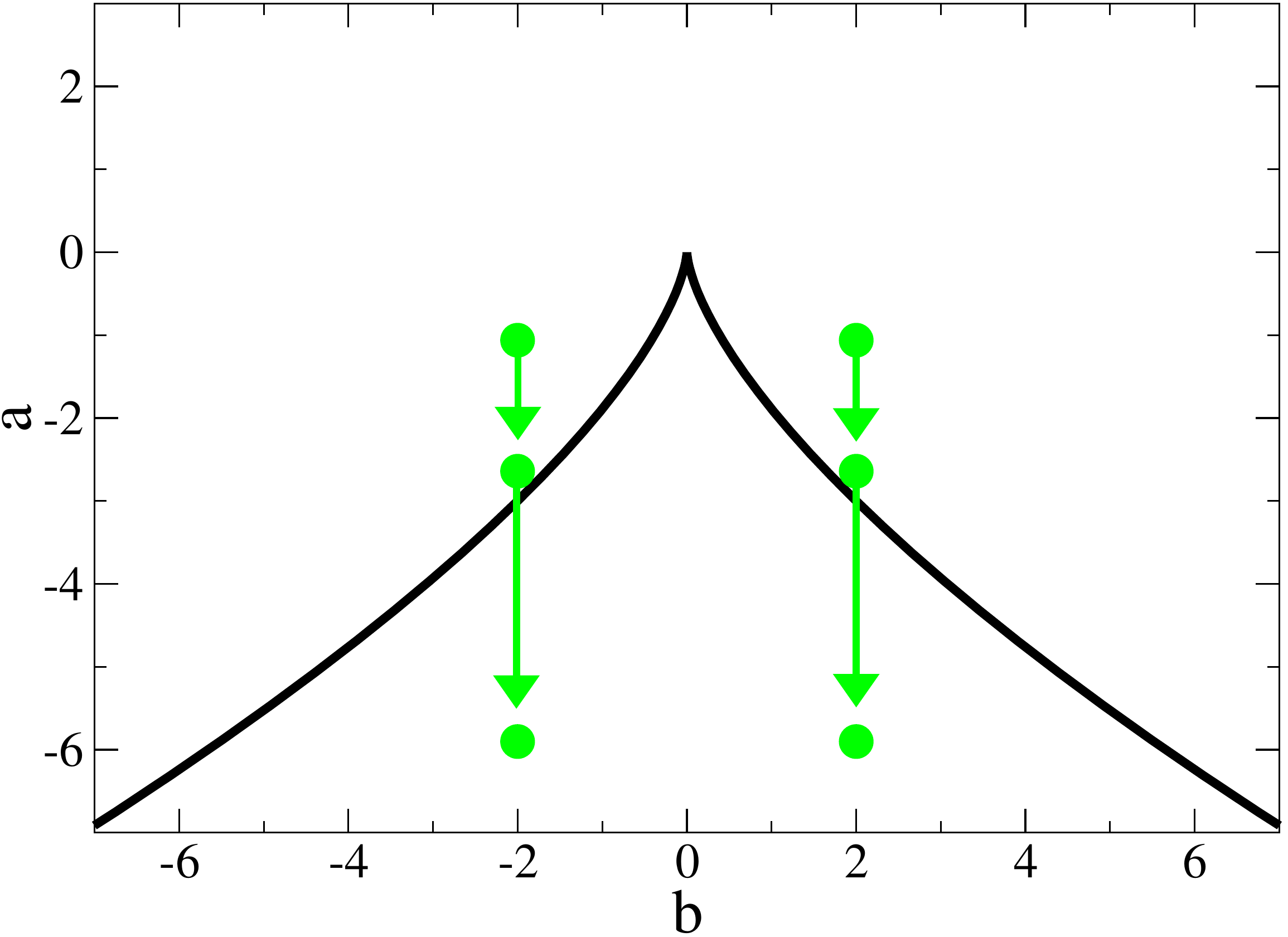}}
\caption{Fixed point stability is determined by the coordinates 
  $(a,b)=(-A,-2x_{f})$ in the $A_3$ catastrophe control parameter space.  They are represented by the green circles.  (b) The fixed points move along the $b$ axis as $R$ is varied.  Since they remain outside the cusp region for all $R$, they will always have focal stability.  (c) The fixed points move along the $a$ axis as $A$ is varied.  The transition from outside to inside the cusp forces a transition in stability from focal to saddle.  The governing equation is (\ref{eq:3dode}) using $G_1(x_1;R)=x^2-R$.}
\label{fig:A3}
\end{minipage}
\end{figure}

A pair of symmetry related fixed points are created 
in a saddle-node bifurcation as $R$ passes through zero.  
Their stability can be determined directly 
by evaluating the zeros of (\ref{eq:cpn3}).  
A more convenient method is 
to examine the evolution of the fixed points in the 
catastrophe control parameter space $(a,b)=(-A,-2x_{f})$.  
This method allows us to predict the fixed point stability 
as the control parameters are changed.  
The fixed point stability for $R<0$ and $R=1$ is 
plotted in Fig. \ref{fig:A3}(b) using green dots.  Their 
evolution as a function of $R$ is indicated 
by the green arrows.

\begin{figure}[htbp] %%%   Fig. 3
\begin{minipage}{\columnwidth}
\centering
\subfigure[$R$ = -1]{
  \includegraphics[height=6cm,width=\columnwidth]{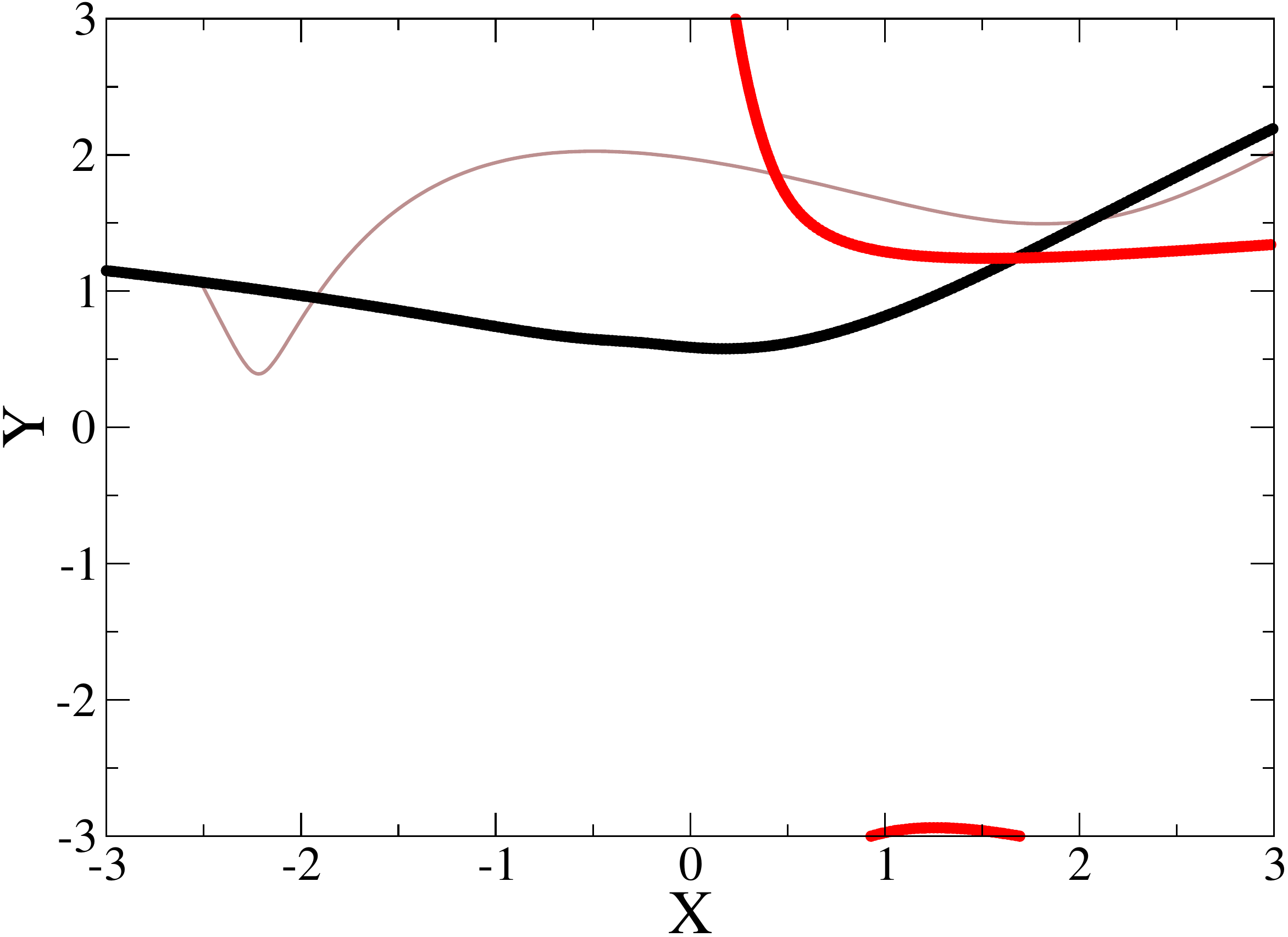}}
\subfigure[$R$ = 0]{\includegraphics[height=6cm,width=\columnwidth]{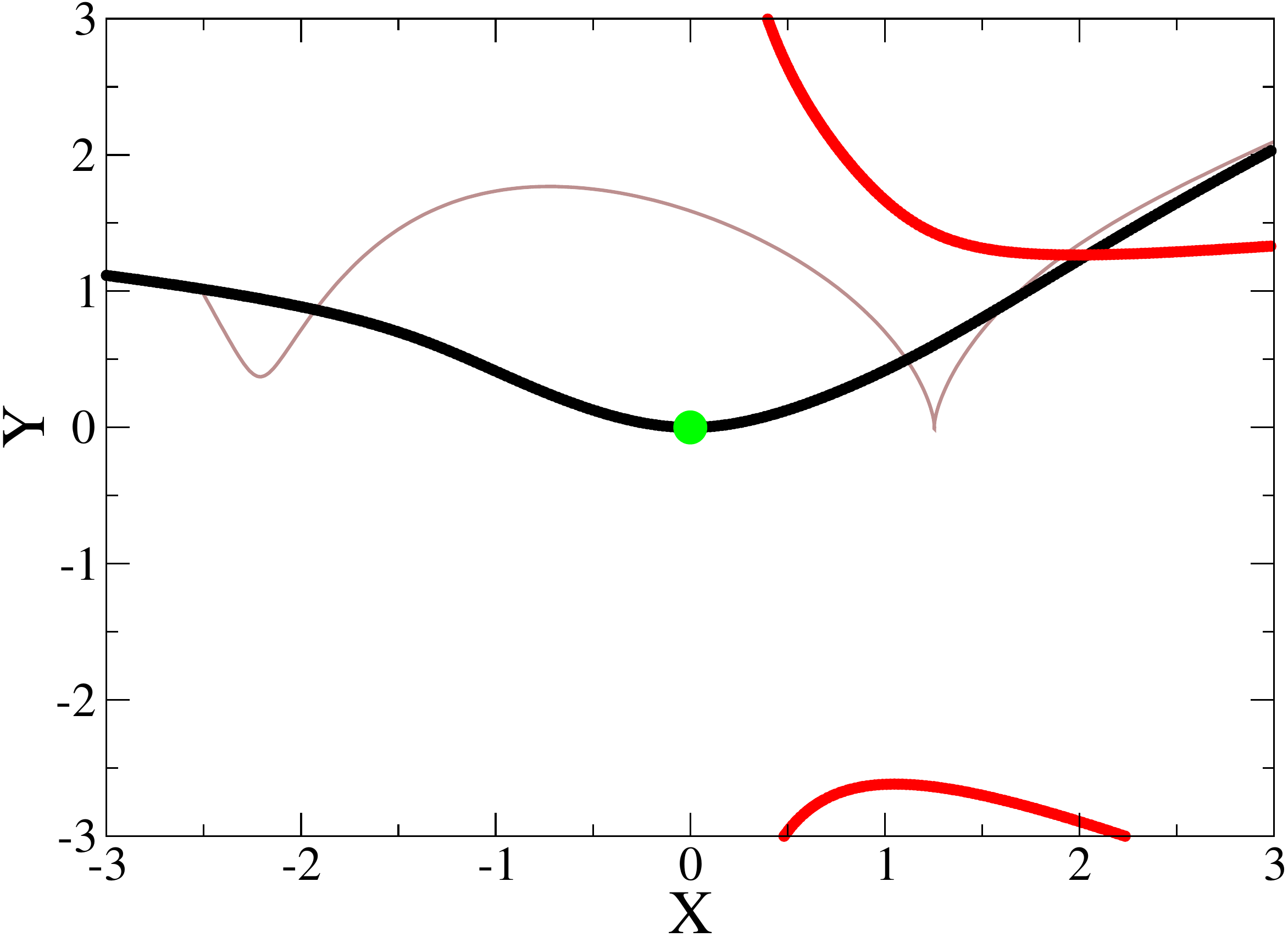}}
\subfigure[$R$ = 1]{\includegraphics[height=6cm,width=\columnwidth]{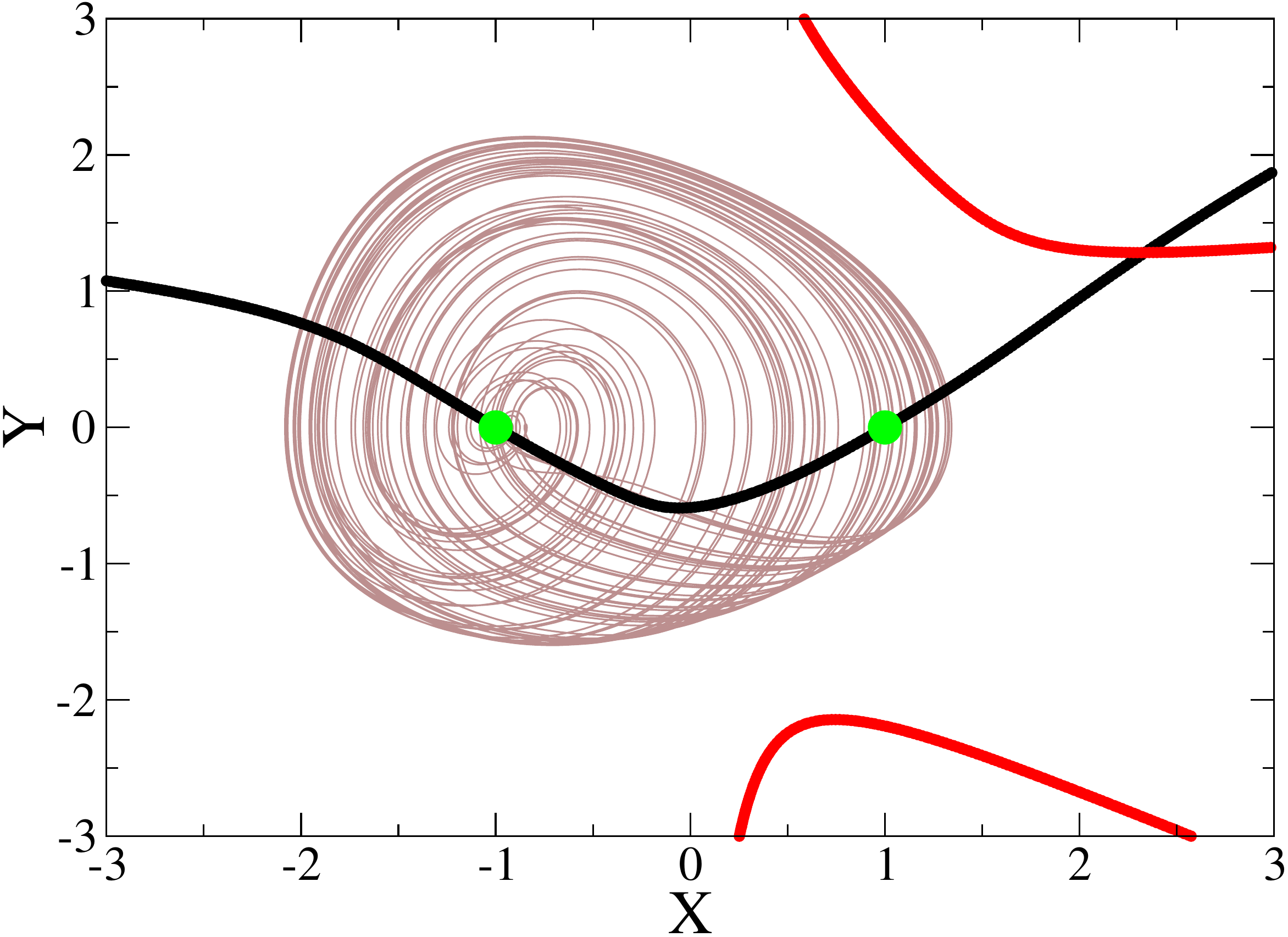}}
\caption{The connecting curves, fixed points and phase space trajectories are 
  plotted in the X-Y plane for $R=(-1,0,1)$.  The governing equation is (\ref{eq:3dode}) 
  using $G_1(x_1;R)=x^2-R$ with parameter values $(A,B)=(-1.7,-0.8)$.  Fixed points 
  undergo a saddle-node bifurcation as $R$ passes through zero. (a-b) Trajectories 
  initialized near the vortex core line escape to infinity.  (c) The alternating 
  stability of the two fixed points creates conditions that generate a strange 
  attractor.  }
\label{fig:SaddleNode}  
\end{minipage}
\end{figure}

No real fixed points exist for $R=-1$.  Figure \ref{fig:SaddleNode}(a) 
shows that vortex and strain curves exist in the phase space despite the 
lack of fixed points.  Phase space trajectories initialized near the core 
curve spiral along its length towards the right.  These trajectories are 
unbounded.  

%%%%%%%%%%%%%%%%%%%%%%%%%%%%%%%%%%%%%%%%%%%%%%%%%
%%%%%%%%%%%%%%%%%%%%%%%%%%%%%%%%%%%%%%%%%%%%%%%%%

A real degenerate fixed point is created when $R=0$.  It has 
coordinates $(a,b)=(1.7,0)$ in the catastrophe control 
parameter plane.  Figure \ref{fig:A3}(b) shows that it is 
located directly above the cusp along the symmetry axis.  
Because it is both outside the cusp and on the symmetry 
axis, the fixed point is a center with a stable inset on 
the left and a unstable outset on the right.  It appears 
on the vortex core curve in Fig. \ref{fig:SaddleNode}(b).  
Phase space trajectories spiral along the core curve towards 
the right and remain unbounded.  

Two real fixed points are created as $R$ is increased past zero.  
As seen in Fig. \ref{fig:A3}(b), they move horizontally along the 
$b$ axis as $R$ is increased.  Since the fixed point stability evolves 
outside of the cusp region for all $R \geq 0$, the fixed points always 
have focal stability for the current value of $A$.  When $R=1$, 
the alternation of (stable focus - unstable outset) with 
(unstable focus - stable inset) builds up the strange attractor 
shown in Fig. \ref{fig:SaddleNode}(c).

%%%%%%%%%%%%%%%%%%%%%%%%%%%%%%%%%%%%%%%%%%%%%%%%%
%%%%%%%%%%%%%%%%%%%%%%%%%%%%%%%%%%%%%%%%%%%%%%%%%

\begin{figure}[htbp] %%%   Fig. 4
\begin{minipage}{\columnwidth}
\centering
\subfigure[$R$ = 1.5385]{
  \includegraphics[height=6cm,width=\columnwidth]{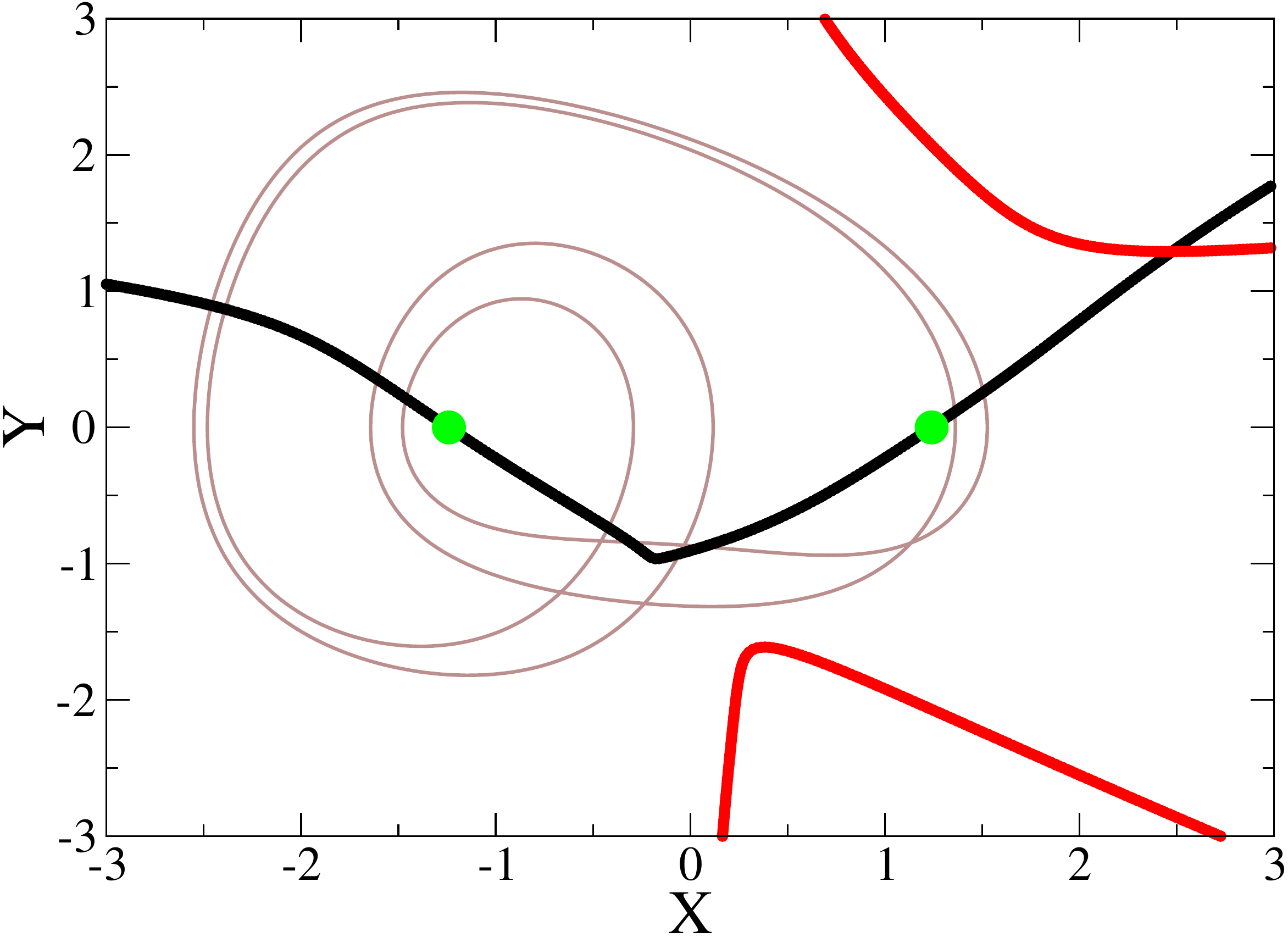}}
\subfigure[$R$ = 1.6264]{\includegraphics[height=6cm,width=\columnwidth]{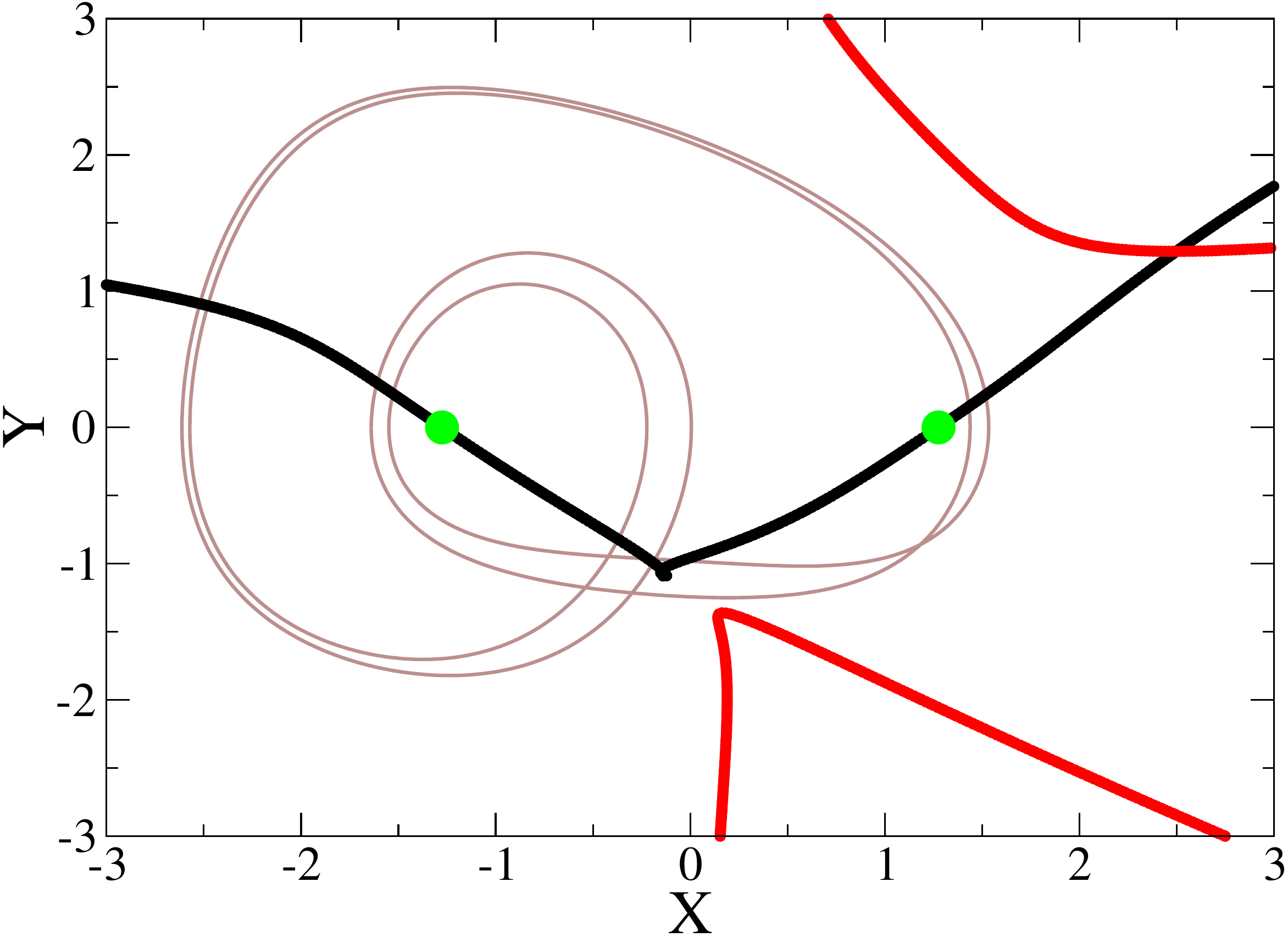}}
\subfigure[$R$ = 1.7143]{\includegraphics[height=6cm,width=\columnwidth]{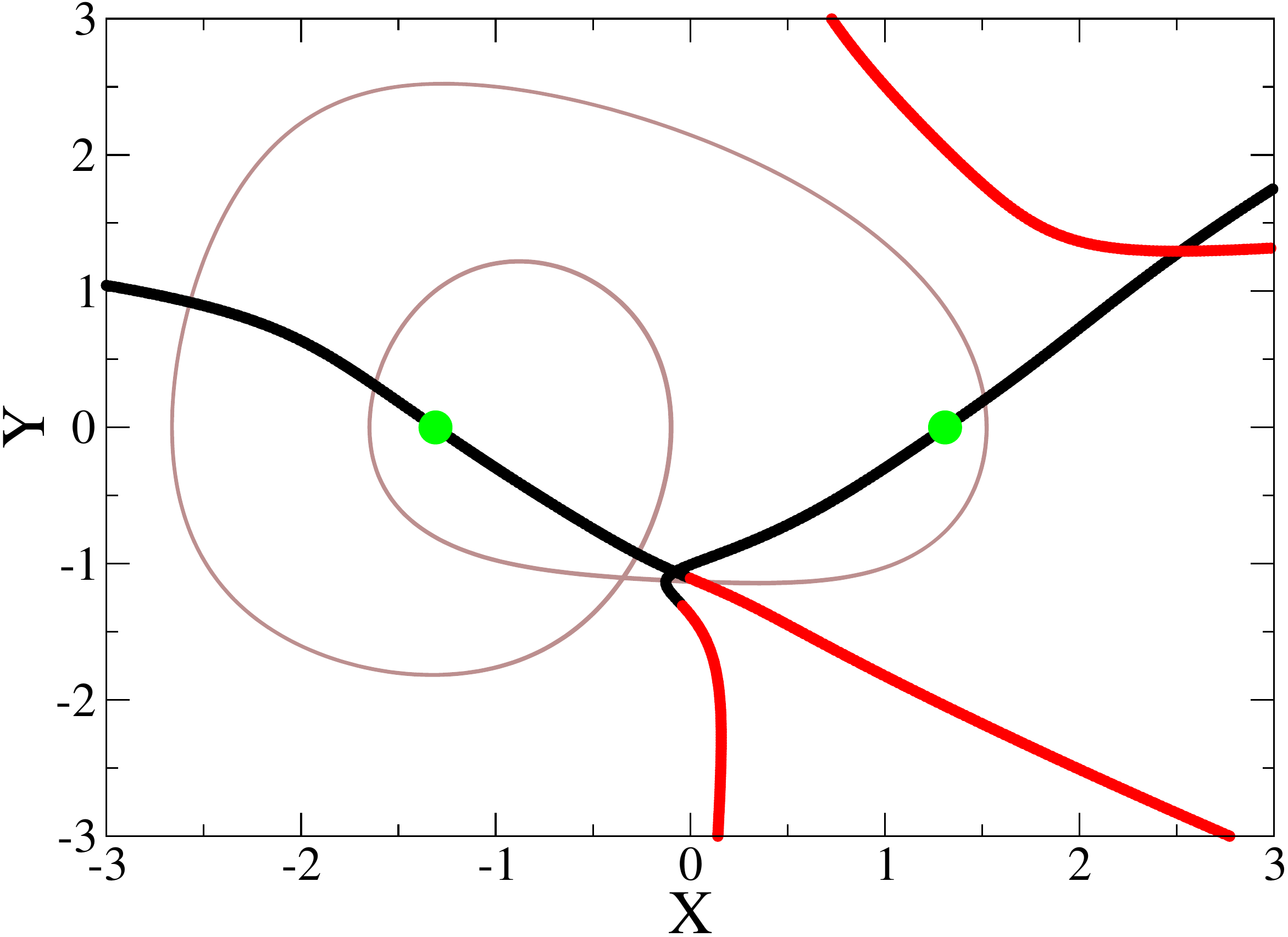}}
\caption{A reconnection between a vortex core curve and a strain curve is shown in the X-Y plane as $R$ is increased past one.  The stable period four and period two orbits are shown in brown.  The governing equation is (\ref{eq:3dode}) using $G_1(x_1;R)=x^2-R$ with parameter values $(A,B)=(-1.7,-0.8)$.}
  \label{fig:reconnect} 
\end{minipage}
\end{figure}

A reconnection between two connecting curves takes place as $R$ is 
increased past one.  The sequence is shown 
in Fig. \ref{fig:reconnect}.  The vortex core curve running between 
the two fixed points breaks and re-attaches itself with the strain 
curve approaching from the bottom.  This sequence is important because 
it indicates that reconnections can be made between vortex and strain 
curves despite their different stability properties.

%%%%%%%%%%%%%%%%%%%%%%%%%%%%%%%%%%%%%%%%%%%%%%%%%
%%%%%%%%%%%%%%%%%%%%%%%%%%%%%%%%%%%%%%%%%%%%%%%%%

\begin{figure}[htbp] %%%   Fig. 5
\begin{minipage}{\columnwidth}
\centering
\subfigure[$A$ = 1.0617]{
  \includegraphics[height=6cm,width=\columnwidth]{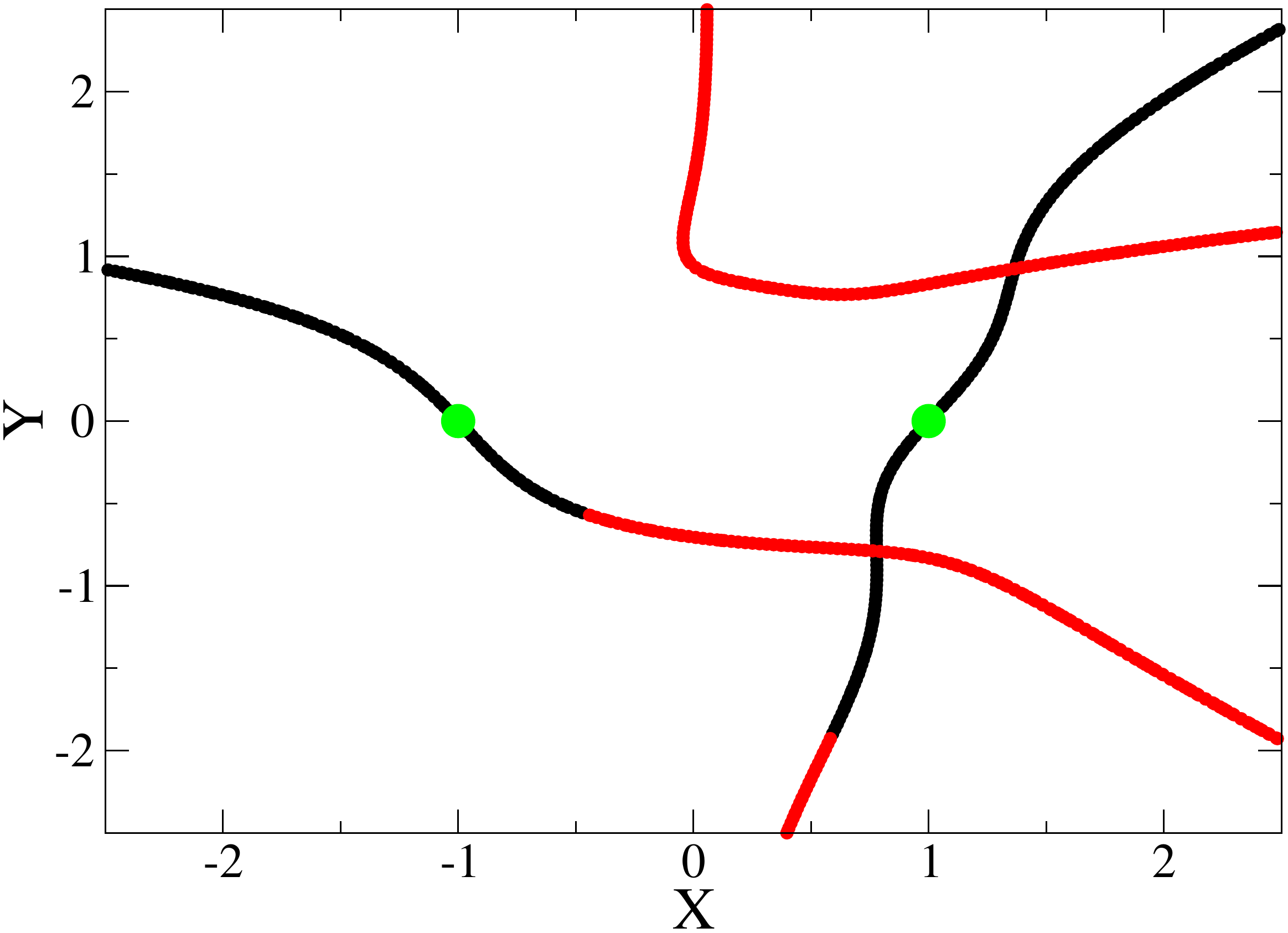}}
\subfigure[$A$ = 2.642]{\includegraphics[height=6cm,width=\columnwidth]{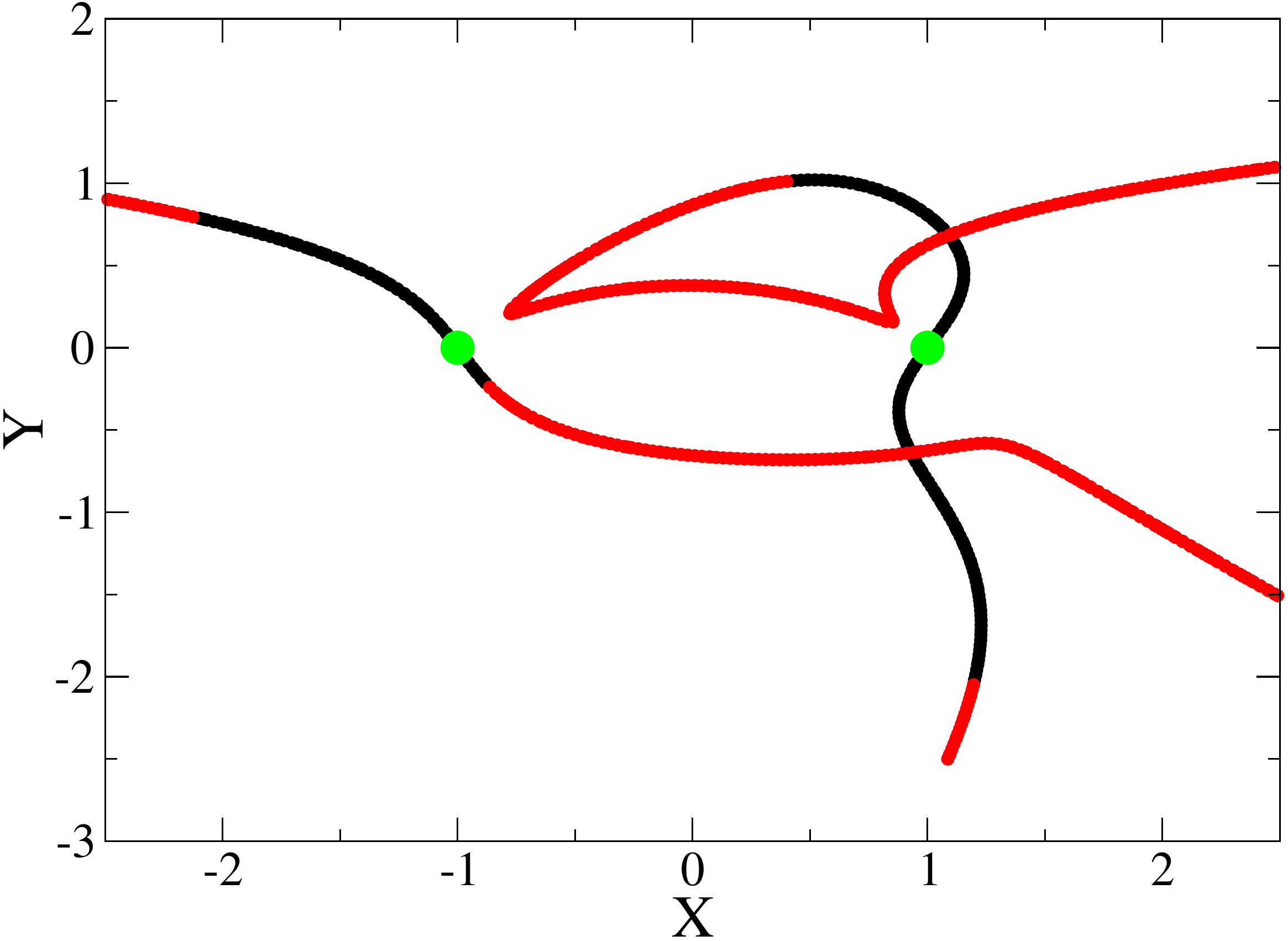}}
\subfigure[$A$ = 5.9012]{\includegraphics[height=6cm,width=\columnwidth]{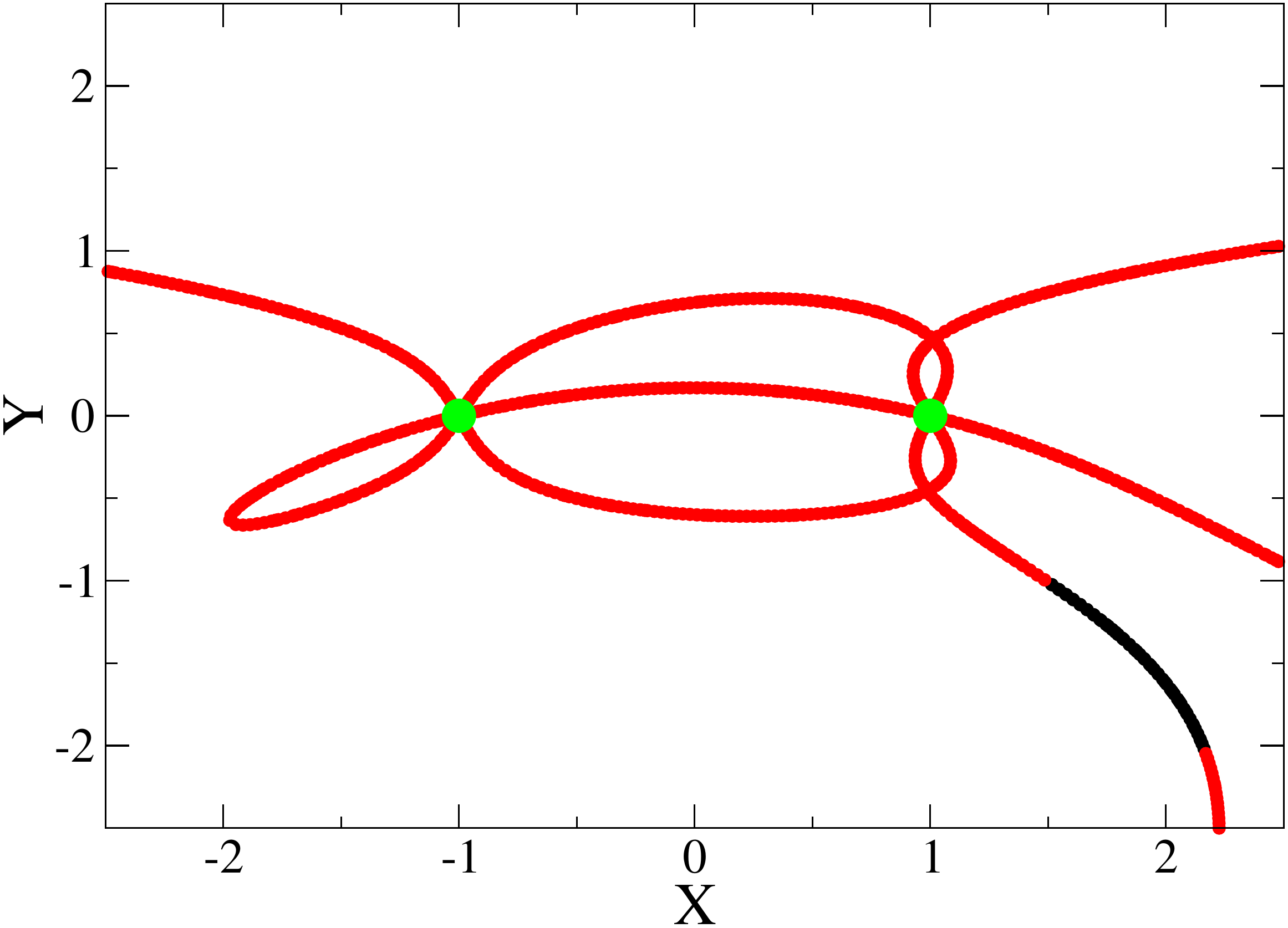}}
\caption{The connecting curve evolution is plotted in the X-Y plane as the parameter $A$ is varied.  The changing stability of the fixed points force local changes in the stability of the connecting curves.  The governing equation is (\ref{eq:3dode}) using $G_1(x_1;R)=x^2-R$ with parameter values $(B,R)=(-0.8,1)$.}
  \label{fig:VortStrain} 
\end{minipage}
\end{figure}

We end this sub-section by fixing the control 
parameters $(B,R)=(-0.8,1)$ and varying $A$.  
The fixed points now move along the 
$a$ axis and transition from outside to inside the 
cusp region.  This transition is illustrated in 
Fig. \ref{fig:A3}(c) for three different values of $A$.  
The change from focal to saddle stability forces a corresponding 
change in stability along the connecting curve near 
the fixed point.  When $a<-3$, all the eigenvalues become 
real and only strain curves can pass 
through the fixed points.  The sequence in 
Fig. \ref{fig:VortStrain} shows the connecting 
curves during this transition.

%%%%%%%%%%%%%%%%%%%%%%%%%%%%%%%%%%%%%%%%%%%%%%%%%
%%%%%%%%%%%%%%%%%%%%%%%%%%%%%%%%%%%%%%%%%%%%%%%%%

\subsection{$m=3$}

\begin{figure}[htbp] %%%   Fig. 6
\begin{minipage}{\columnwidth}
\center
\includegraphics[height=6cm,width=\columnwidth]{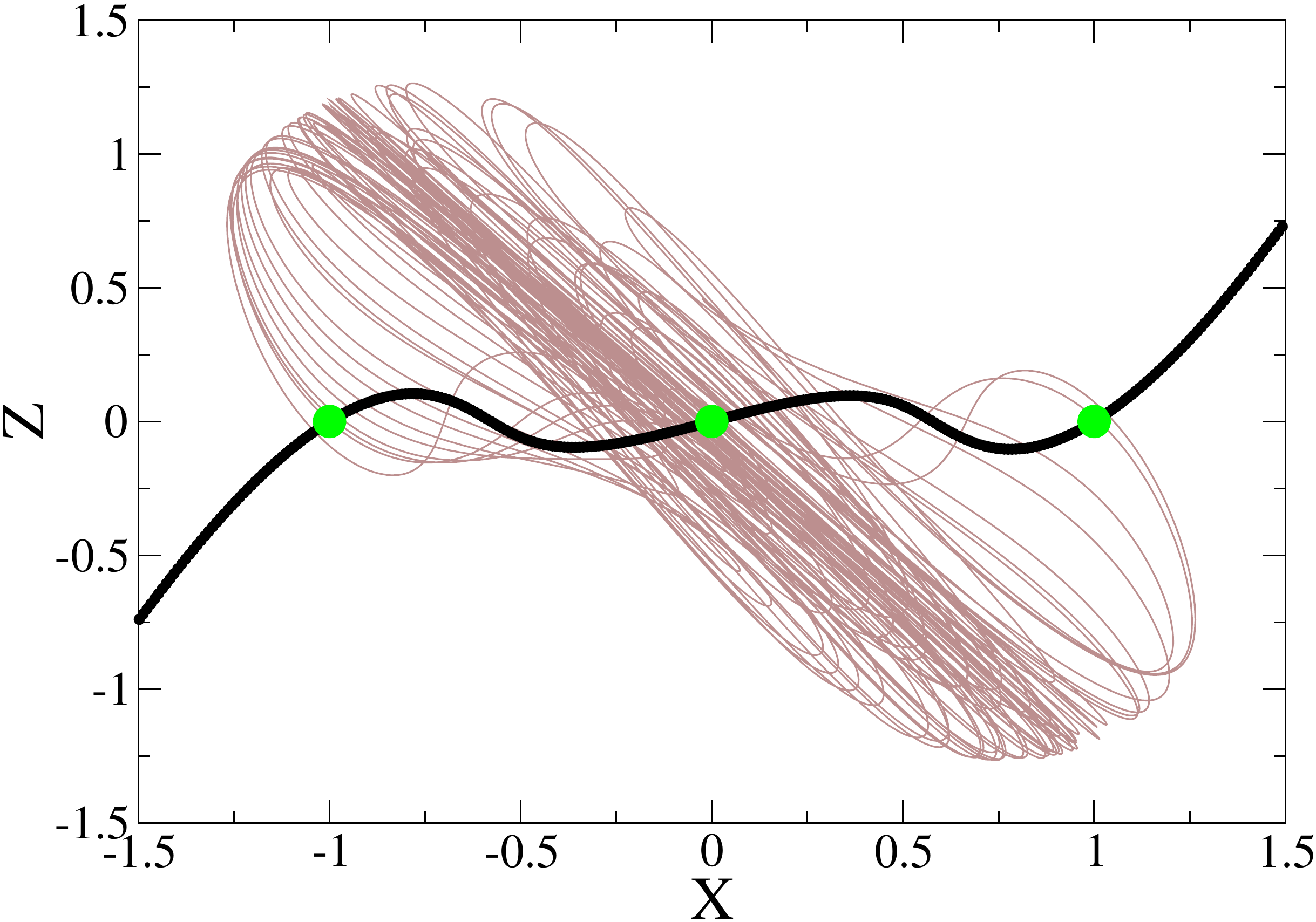}
\caption{X-Z projection of the strange attractor generated by 
Eq. (\ref{eq:3dode}) with $G_1(x,R)=x^3-Rx$.  The flow is organized 
by the fixed points and the vortex core curve.  The fixed point 
at the origin is a center with stable insets.  The symmetric fixed 
points are foci with unstable outsets.  Parameter values: $(A,B,R)=(-1.25,-0.8,1)$.}
\label{fig:3d3fpXY} 
\end{minipage}
\end{figure}

For $G_1(x_1;R)=x^3-Rx$, the jacobian is given by Eq. (\ref{eq:jcn3}) with 
$G_1'(x_1;R) = 3x^2-R$.  The description of the fixed point 
stability proceeds as in the case with $m=2$.
The fixed point at the origin has coordinates 
$(a,b)=(-A,0)$ in the catastrophe control parameter plane.  
Because it sits on the symmetry axis, the fixed point is a 
center when $A<0$.  For $A\geq0$, it is a saddle.  
The symmetry related fixed points 
have the same coordinate $(a,b)=(-A,-2R)$ in the catastrophe 
control parameter plane.  They have three real eigenvalues 
for $(a/3)^{3}<-R^2$.  Otherwise, the fixed points have one real 
eigenvalue and a complex conjugate pair.  Figure \ref{fig:3d3fpXY} 
shows the strange attractor, connecting curves and fixed points 
generated for parameter values $(A,B,R)=(-1.25,-0.8,1)$.  The 
fixed point at the origin has two stable insets that attract 
flow along the vortex core curve from the two outer fixed points.

%%%%%%%%%%%%%%%%%%%%%%%%%%%%%%%%%%%%%%%%%%%%%%%%%
%%%%%%%%%%%%%%%%%%%%%%%%%%%%%%%%%%%%%%%%%%%%%%%%%

\section{Four Dimensions}
\label{sec:4D}

\begin{figure}[htbp] %%%   Fig. 7
\begin{minipage}{\columnwidth}
\centering
\subfigure[Bifurcation set for the swallowtail catastrophe $A_4$.  The number of real eigenvalues in each region are indicated.]{
  \includegraphics[height=5cm,width=\columnwidth]{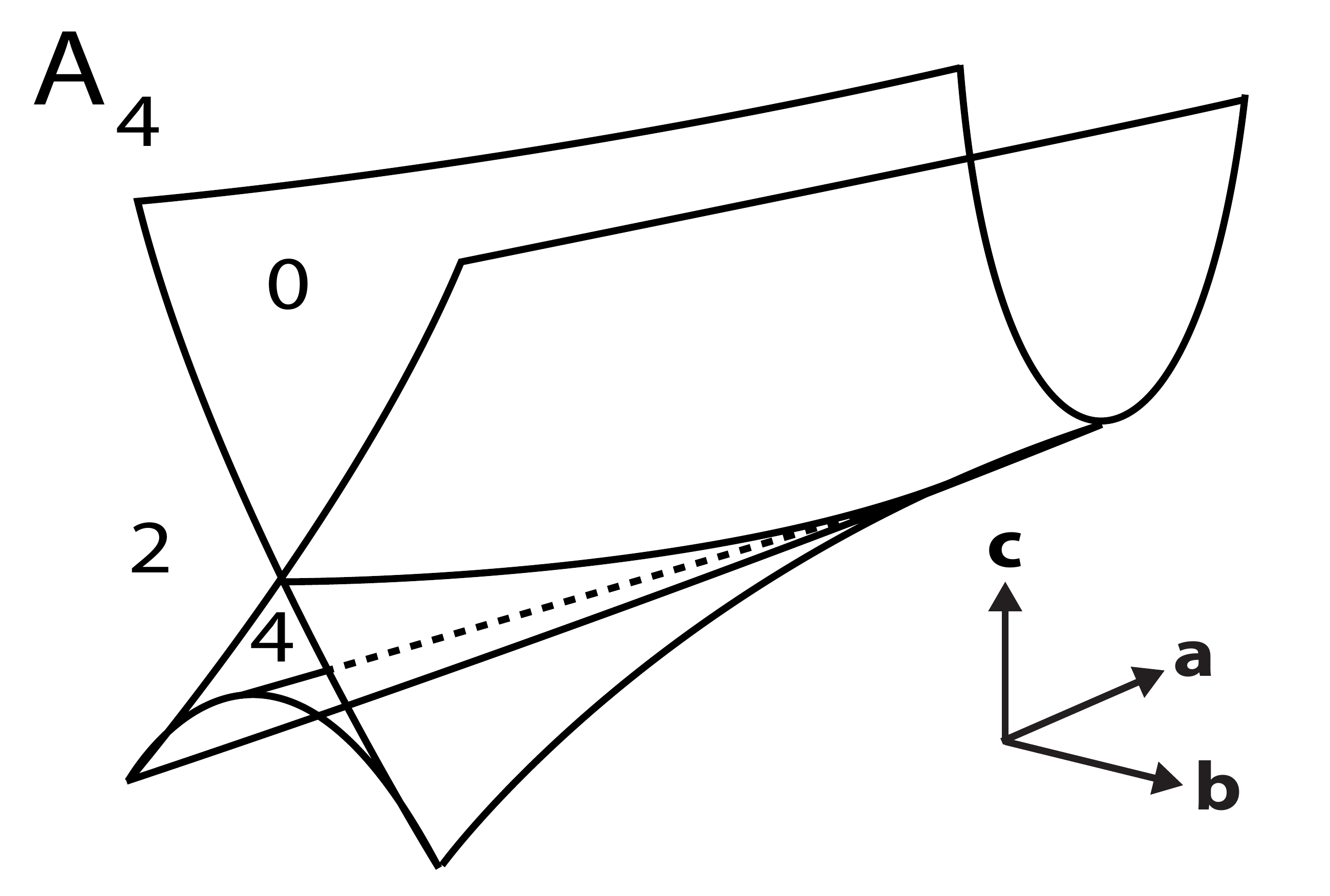}}
\subfigure[Cross section of the bifurcation set taken at $a=-2$.  The fixed point evolution is shown for $(A,B,C)=(0,2,-1)$ and $R=(1,-0.5,-2)$ (bottom to top)]{\includegraphics[height=6cm,width=\columnwidth]{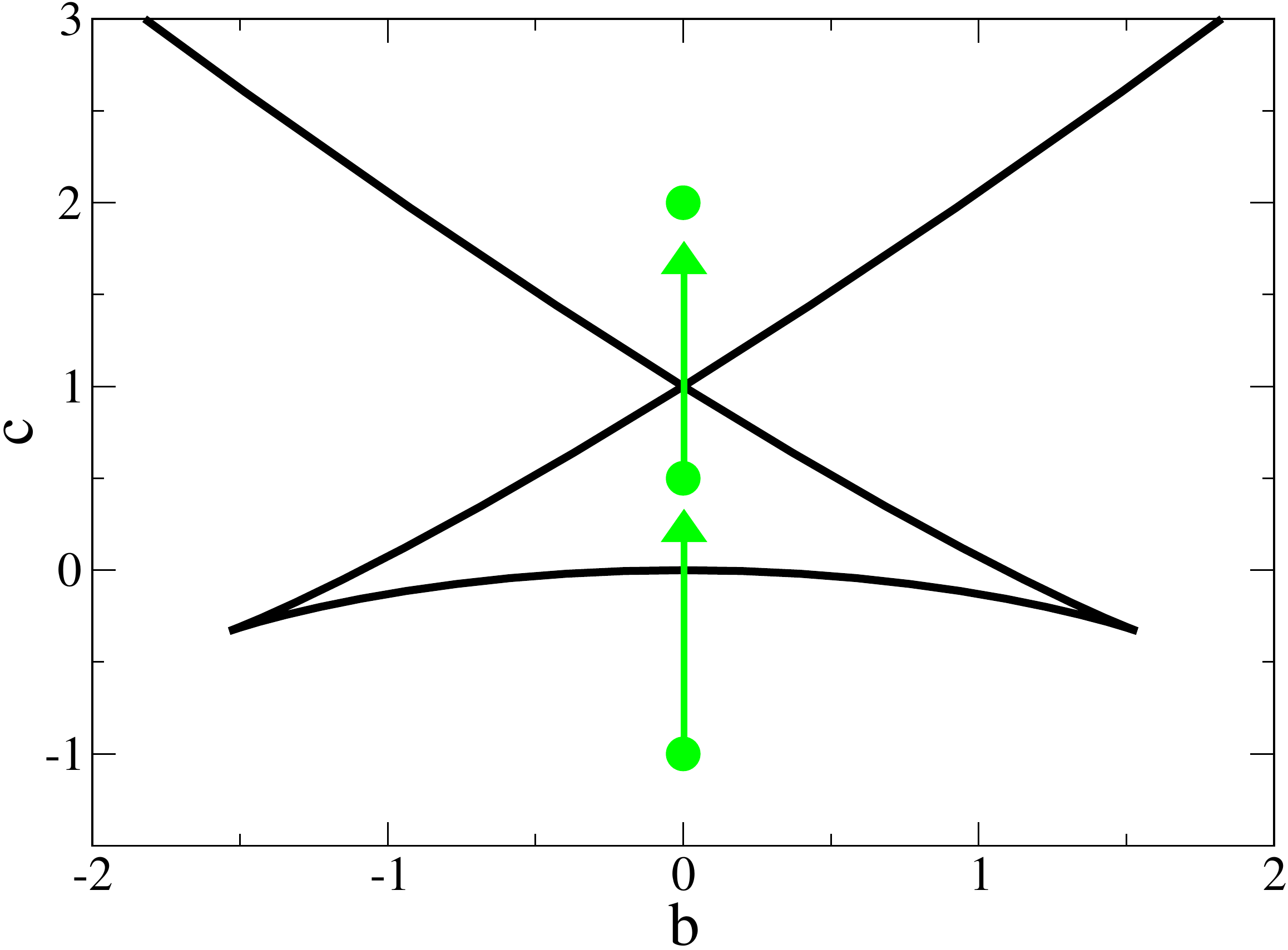}}
\subfigure[Cross section of the bifurcation set taken at $a=2.335$.  The fixed point evolution is shown for $(A,B,C)=(-3.2,-2.335,-1)$ and $R=(0.15,1)$ (center to outer pair)]{\includegraphics[height=6cm,width=\columnwidth]{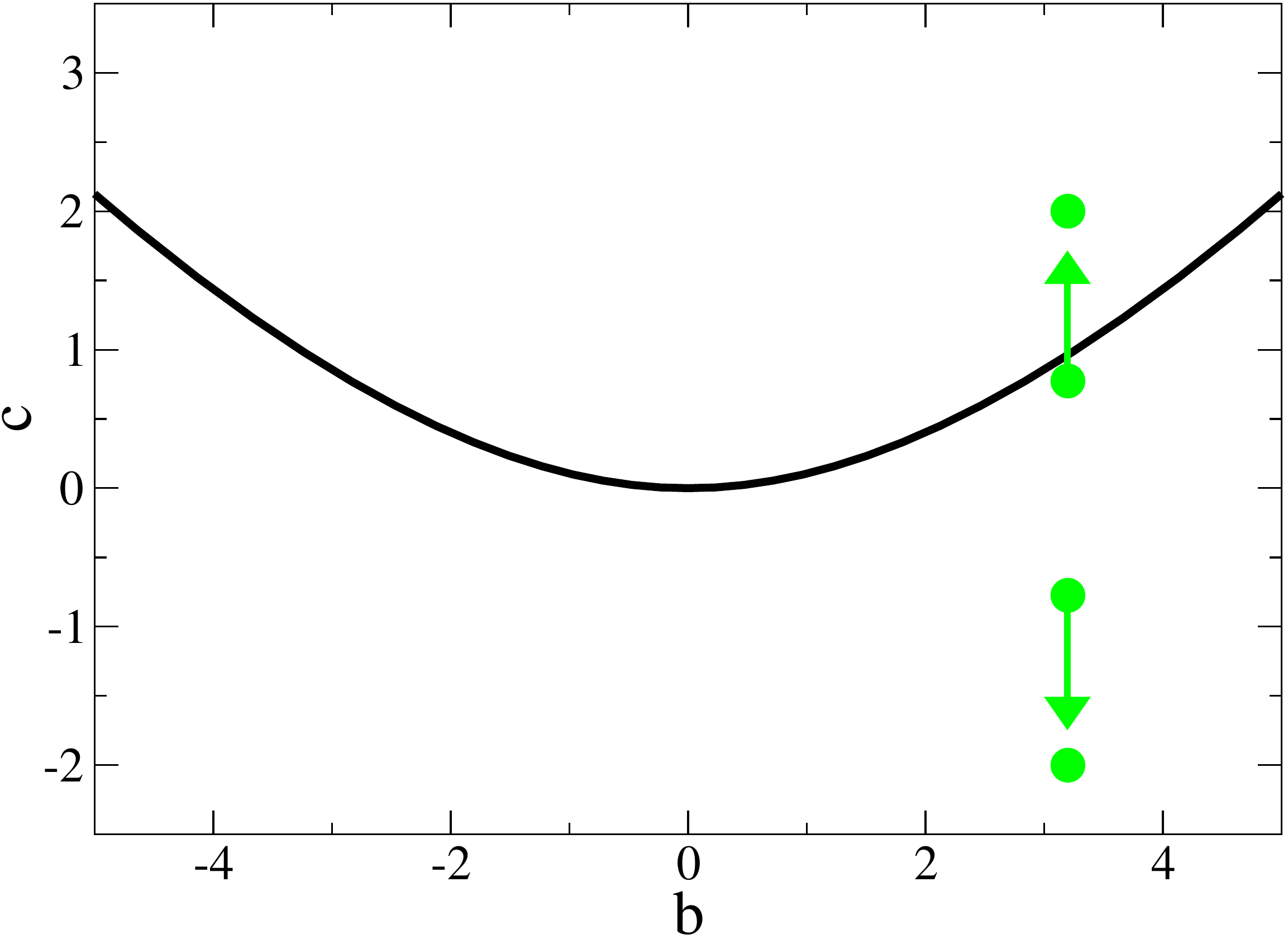}}
\caption{Fixed point stability is determined by the coordinates $(a,b,c)$ in the $A_4$ catastrophe control parameter space.  They are represented by the green circles.  (b) The single fixed point for $G_1(x_1;R)=Rx$ moves through the three regions of stability as $R$ is varied.  (c) The two real fixed point for $G_1(x_1;R)=x^2-R$ move through the two regions of stability as $R$ is varied.}
  \label{fig:A4} 
\end{minipage}
\end{figure}

In this section we set $G_2(y,z,u;A,B,C)=Ay+Bz+Cu^{3}$ in order to 
study four-dimensional differential dynamical systems that take the 
form 

\begin{equation}  \begin{array}{rcl}
  \dot{x} &=& y \\
  \dot{y} &=& z \\
  \dot{z} &=& u \\
  \dot{u} &=& G_{1}(x;R)+Ay+Bz+Cu^{3} \end{array}
\label{eq:4dode}
\end{equation}
The jacobian at any fixed point ${\bf x}_f$ is

\begin{equation}
J = \left[  \begin{array}{cccc}
0 & 1 & 0 & 0 \\ 
0 & 0 & 1 & 0 \\
0 & 0 & 0 & 1 \\
G_1' & A & B & 0 \end{array} \right]_{{\bf x}_{f}}
\label{eq:jcn4}
\end{equation}
with a characteristic polynomial 

\begin{equation}
{\rm det}(J-\lambda I_4))=A_4(\lambda)=\lambda^4-B\lambda^2-A
\lambda - G_1'(x_f;R)
\label{eq:cpn4}
\end{equation}
whose roots determine the stability 
properties of the fixed point.  By making the 
substitution $(a,b,c)=(-B,-A,-G_1'(x_{f};R))$, 
(\ref{eq:cpn4}) becomes the canonical 
unfolding of the swallowtail catastrophe 
$A_4$ whose bifurcation 
set is shown in Fig. \ref{fig:A4}(a).  
The control parameter space is divided into three disjoint
open regions that describe fixed points with four, two, or
zero real eigenvalues and zero, one, or two pairs of
complex conjugate eigenvalues.  The open regions
are connected and simply connected.  The bifurcation sets
separating these open regions satisfy $A_4(\lambda)=0$
and $dA_4(\lambda)/d\lambda =0$.    

We use polynomials $G_1$ of degree $m=1,2,3$ in the sections below to 
study the properties of connecting curves in higher dimensions.  
For $m=1$, we set $B>0$ and vary the parameter $R$ such that the single 
fixed point traverses the three regions of stability shown in 
Fig. \ref{fig:A4}(b).  We describe the effects on the connecting curves 
as the fixed point passes through each of the regions.  We also determine 
the global stability of the connecting curve as a function of the single 
phase space variable $u$.  

For $m=2,3$ we set $B<0$ such that the control parameter space is 
divided into two regions where (\ref{eq:jcn4}) produces one or two 
pairs of complex conjugate eigenvalues.  Under certain conditions, 
the fixed points can assume different values of the 
index $\kappa$.  We describe these effects on the structure 
of strange attractors and their connecting curves.

%%%%%%%%%%%%%%%%%%%%%%%%%%%%%%%%%%%%%%%%%%%%%%%%%
%%%%%%%%%%%%%%%%%%%%%%%%%%%%%%%%%%%%%%%%%%%%%%%%%

\subsection{$m=1$}

To investigating the global stability 
of connecting curves change in higher 
dimensions, we fix $(A,B,C)=(0,2,-1)$ and vary $R$ for 
$G_1(x_1;R)=Rx$.  The stability of the single fixed 
point at the origin depends on its coordinates 
$(a,b,c)=(-2,0,-R)$ in the catastrophe control 
parameter space.  As $R$ is varied, the fixed point 
moves along the symmetry axis $(b=0)$ and passes 
through the three regions of stability shown in 
Fig. \ref{fig:A4}(b).  Along the symmetry 
axis, the fixed point has the following eigenvalue 
spectrum: one degenerate complex conjugate eigenvalue pair when 
$c<0$; none for $0<c<(a/2)^2$; and a doubly degenerate 
pair for $c>(a/2)^2$.  

Although useful, the fixed points only provide limited 
(local) information about the stability of a connecting 
curve.  The stability of an arbitrary point in phase space 
is determined by the characteristic equation of the 
jacobian evaluated at that point.  For the control 
parameters listed above, the characteristic equation 
is 

\begin{equation} 
{\rm det}(J-\lambda I_4))=
C_p(\lambda,u;R)=\lambda^4+3 u^2 \lambda^3-2 \lambda^2 -R
\label{eq:cpglobal}
\end{equation}
which is a function of the single phase space variable $u$.  
Equation (\ref{eq:cpglobal}) is used to create a simple 
partition of the phase space that determines the stability 
along the entire length of the connecting curve.  

\begin{figure}[htbp] %%%   Fig. 8
\begin{minipage}{\columnwidth}
\centering
\subfigure[$R = 1$]{
  \includegraphics[height=6cm,width=\columnwidth]{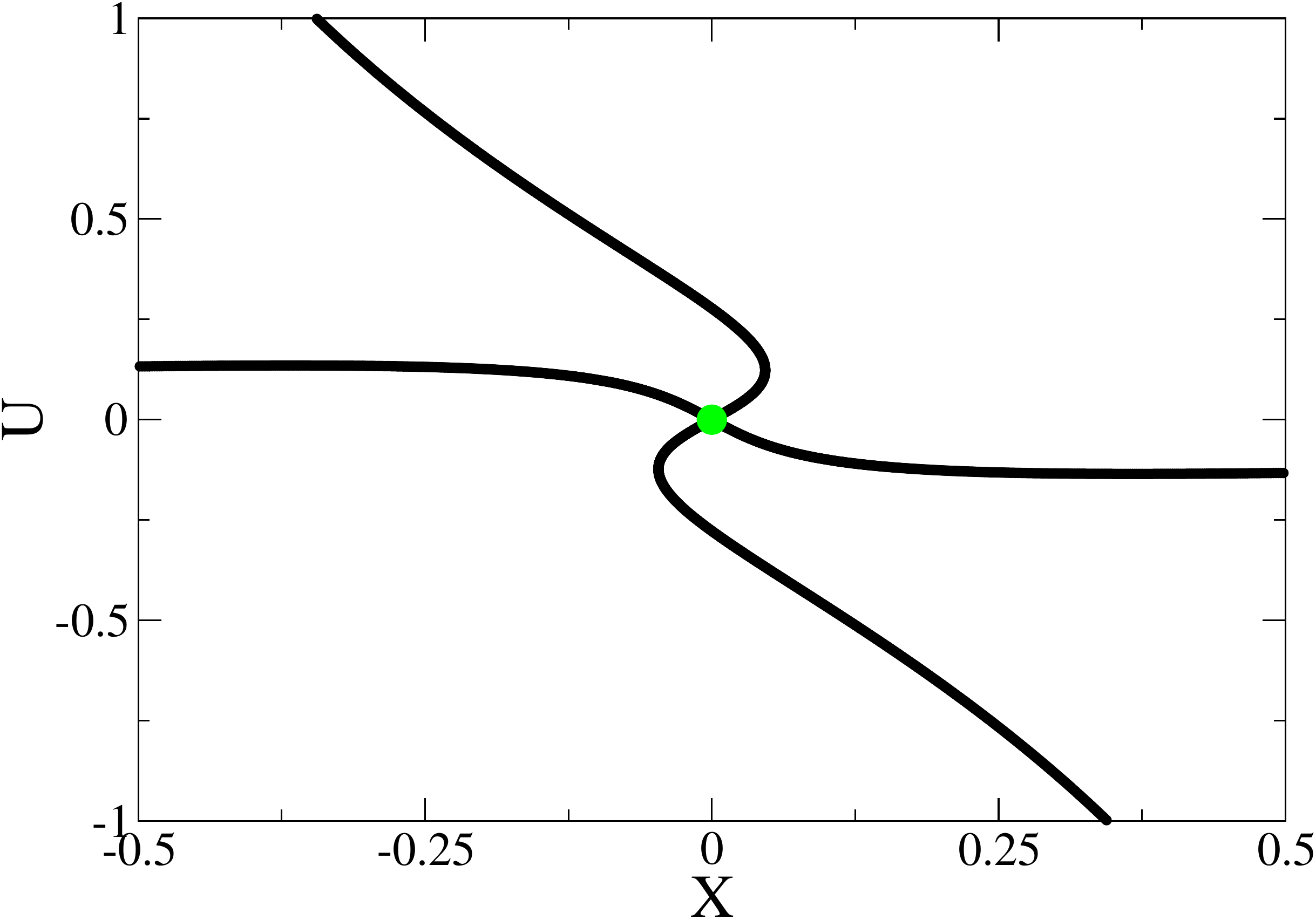}}
\subfigure[$R = -0.5$]{\includegraphics[height=6cm,width=\columnwidth]{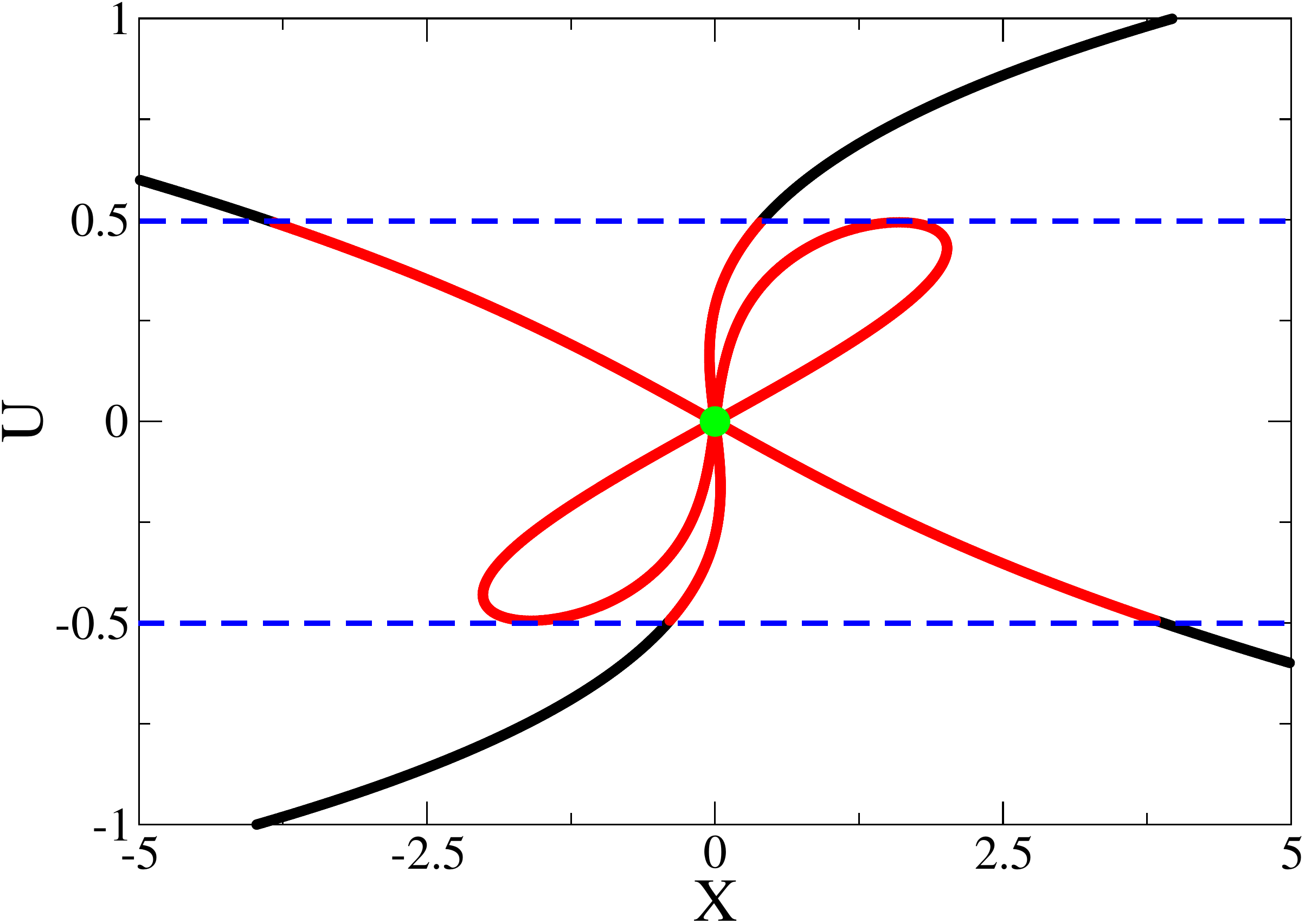}}
\subfigure[$R = -2$]{\includegraphics[height=6cm,width=\columnwidth]{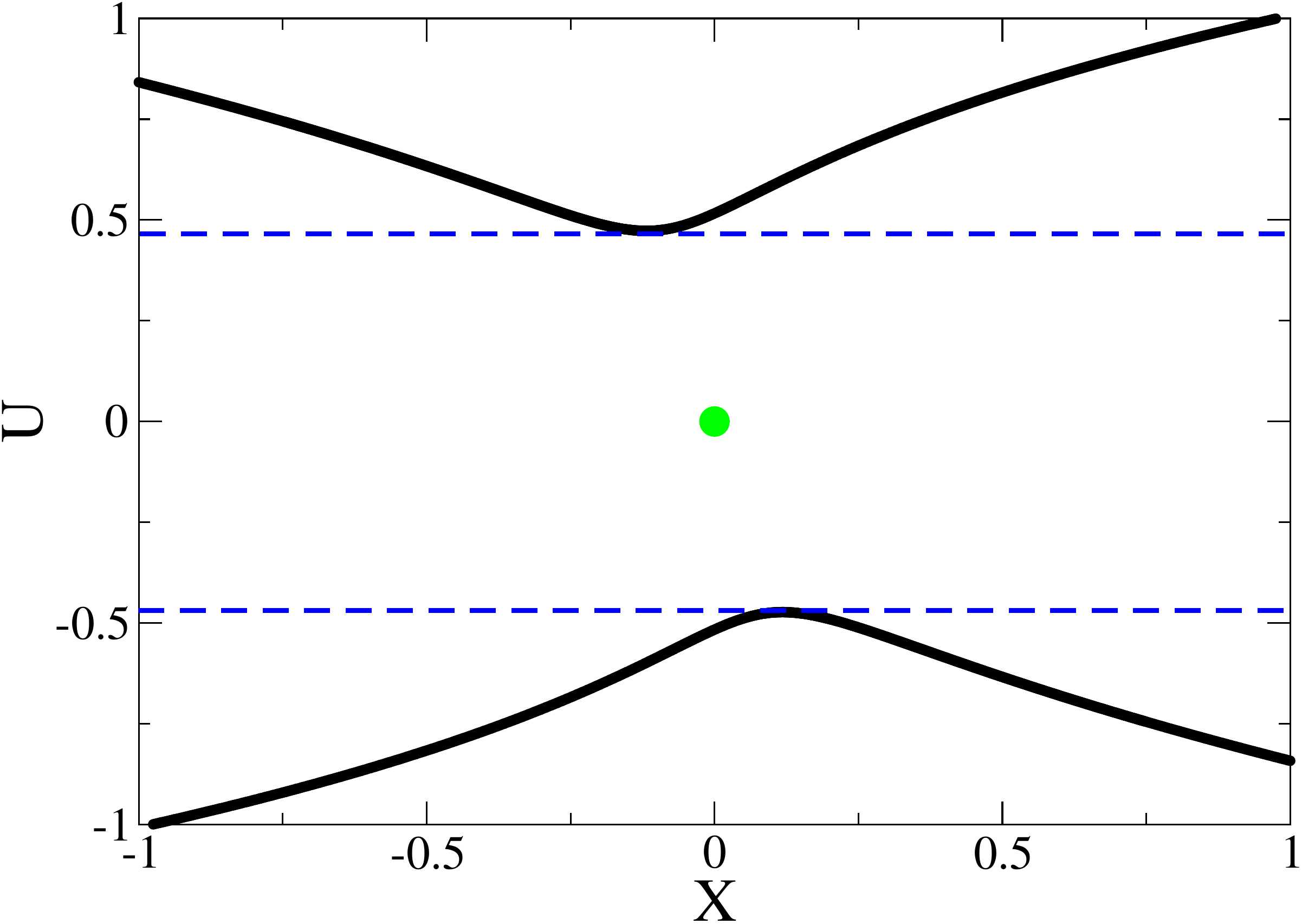}}
\caption{Connecting curves and fixed point from Eq. (\ref{eq:4dode}) are plotted in the $X-U$ plane using $G_1(x,R)=Rx$ with parameter values $(A,B,C)=(0,2,-1)$.  Equation (\ref{eq:cpglobal}) is used to partition the phase space (blue dotted line).  These paritions define the stability of the connecting curves within them.} 
\label{fig:4d1fp}
\end{minipage}
\end{figure}

We start by choosing $R=1$ such that $c<0$.  For this case the 
fixed point has a degenerate complex conjugate eigenvalue pair.  
The index $\kappa=1$.  The $n-2\kappa=2$ vortex core curves that pass 
through the fixed point are shown in Fig. \ref{fig:4d1fp}(a).  
Equation (\ref{eq:cpglobal}) produces a single pair of complex 
conjugate eigenvalues for all $u$.  As a result, the global 
stability properties of the connecting curve remain unchanged 
throughout the phase space.  

Next, we set $R=-0.5$.  This moves the fixed point into 
the region $0<c<(a/2)^2$ where no complex conjugate 
eigenvalue pairs are produced.  Since $\kappa=0$, we expect 
$n-2\kappa=4$ strain curves to connect to the fixed point 
through all possible stable insets and unstable outsets.  The 
strain curves for this value of $R$ are shown in 
Fig. \ref{fig:4d1fp}(b).  Using Eq. (\ref{eq:cpglobal}), the 
phase space is split into two parts.  They are separated by the 
blue dotted line.

%%%%%%%%%%%%%%%%%%%%%%%%%%%%%%%%%%%%%%%%%%%%%%%%%
%%%%%%%%%%%%%%%%%%%%%%%%%%%%%%%%%%%%%%%%%%%%%%%%%

The first part of the phase space (center) contains 
only strain curves because (\ref{eq:cpglobal}) produces 
all real eigenvalues within this range of $u$.  The second part of the phase space 
(top and bottom) contains only vortex core curves because 
(\ref{eq:cpglobal}) produces a single pair of complex conjugate 
eigenvalues.  The abrupt change in stability that is observed to 
take place along the connecting curves is explained by a 
transition between the two phase space partitions.

Setting $R=-2$ moves the fixed point into the final region of the 
catastrophe control parameter space where $c>(a/2)^2$ and $\kappa=2$.  
In this region, $n-2\kappa=0$ connecting curves connect to the fixed point.  
The phase space is again divided into two parts using 
Eq. (\ref{eq:cpglobal}).  The first part (center) produces two pairs 
of complex conjugate eigenvalues.  No vortex core lines are able to enter 
this region.  The second part (outer top and bottom) produces a single pair 
and supports the creation of vortex core lines as shown in 
Fig. \ref{fig:4d1fp}(c).  

\subsection{$m=2$}

\begin{figure}[htbp] %%%   Fig. 9
\begin{minipage}{\columnwidth}
\centering
\includegraphics[height=6cm,width=\columnwidth]{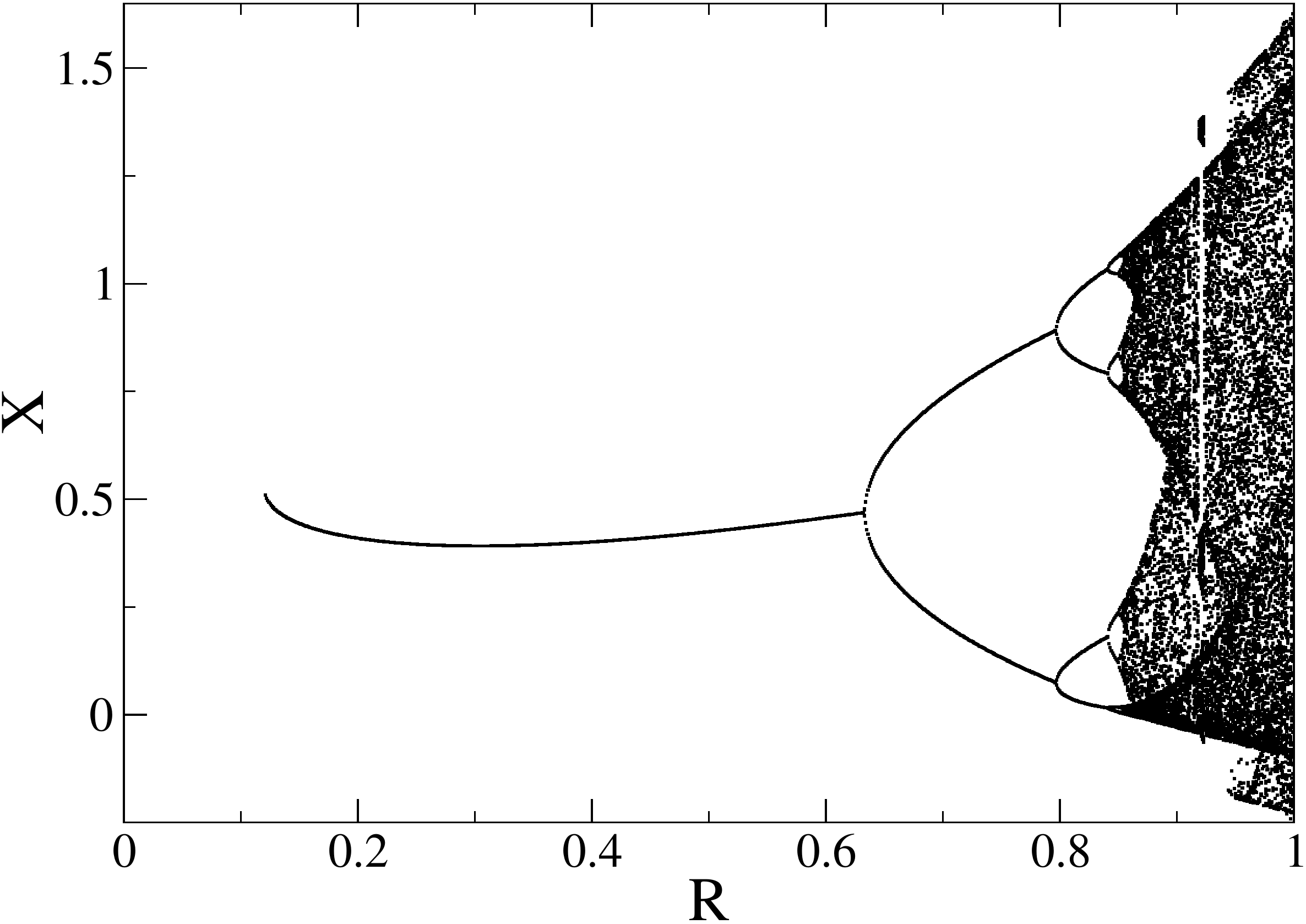}
  \caption{Bifurcation diagram created using a single Poincar\'e surface of 
    section along the X-axis (Y=0) for Eq. (\ref{eq:4dode}) with $G_1(x;R)=x^2-R$.  A period doubling route to chaos is observed.  Parameter values: $(A,B,C)=(-3.2,-2.335,-1)$.}
  \label{fig:4dbif_R} 
\end{minipage}
\end{figure}

For $m=2$ we fix the control parameters $(A,B,C)=(-3.2,-2.335,-1)$ 
and vary $R$ for $G_1(x_1;R)=x^2-R$.  A period doubling route to chaos 
is observed in Fig. \ref{fig:4dbif_R}.  

\begin{figure}[htbp] %%%   Fig. 10
\begin{minipage}{\columnwidth}
\centering
\subfigure[$(A,B,C,R)=(-3.2,-2.335,-1,0.15)$]{
  \includegraphics[height=6cm,width=\columnwidth]{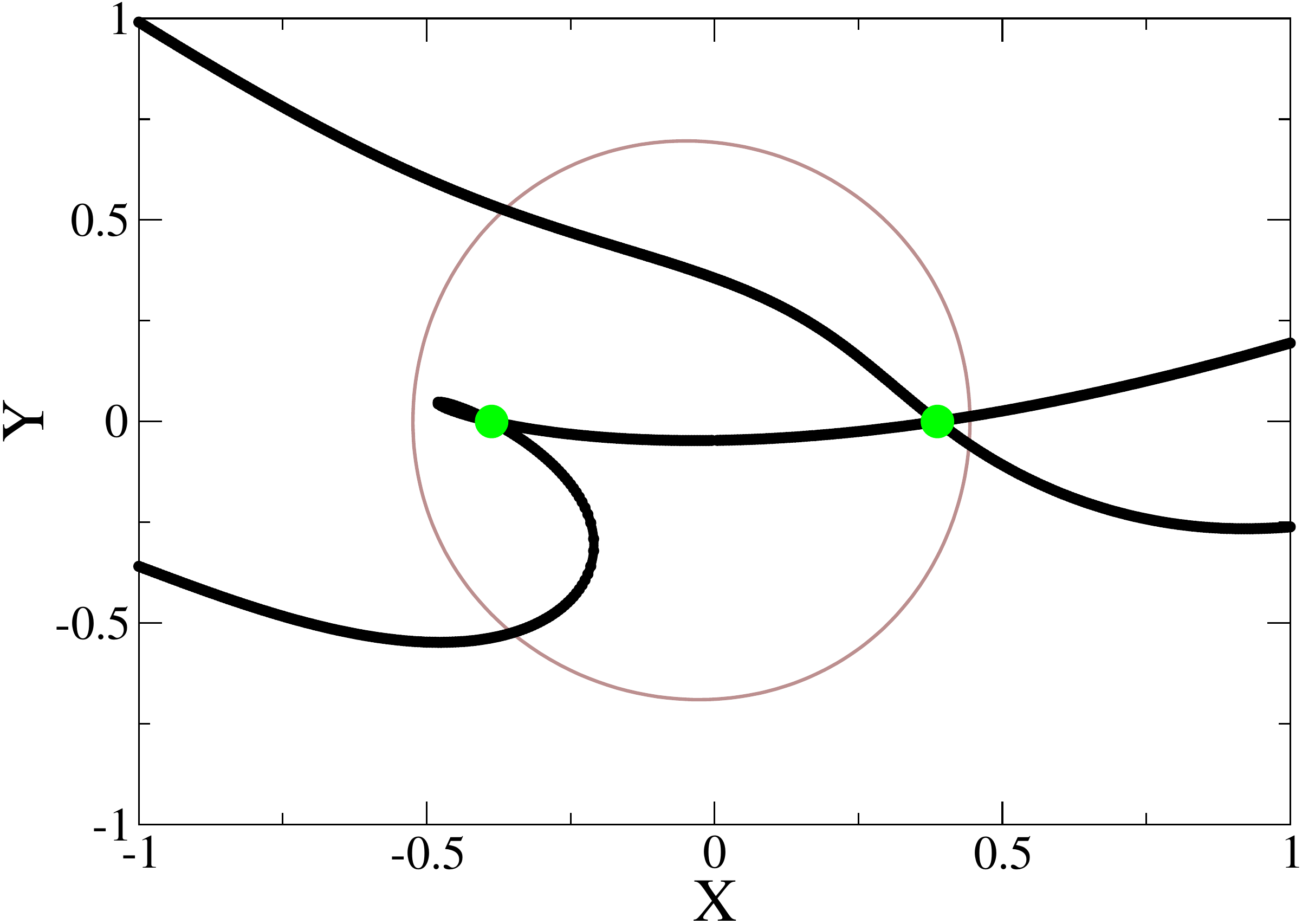}}
\subfigure[$(A,B,C,R)=(-3.2,-2.335,-1,1)$]{
  \includegraphics[height=6cm,width=\columnwidth]{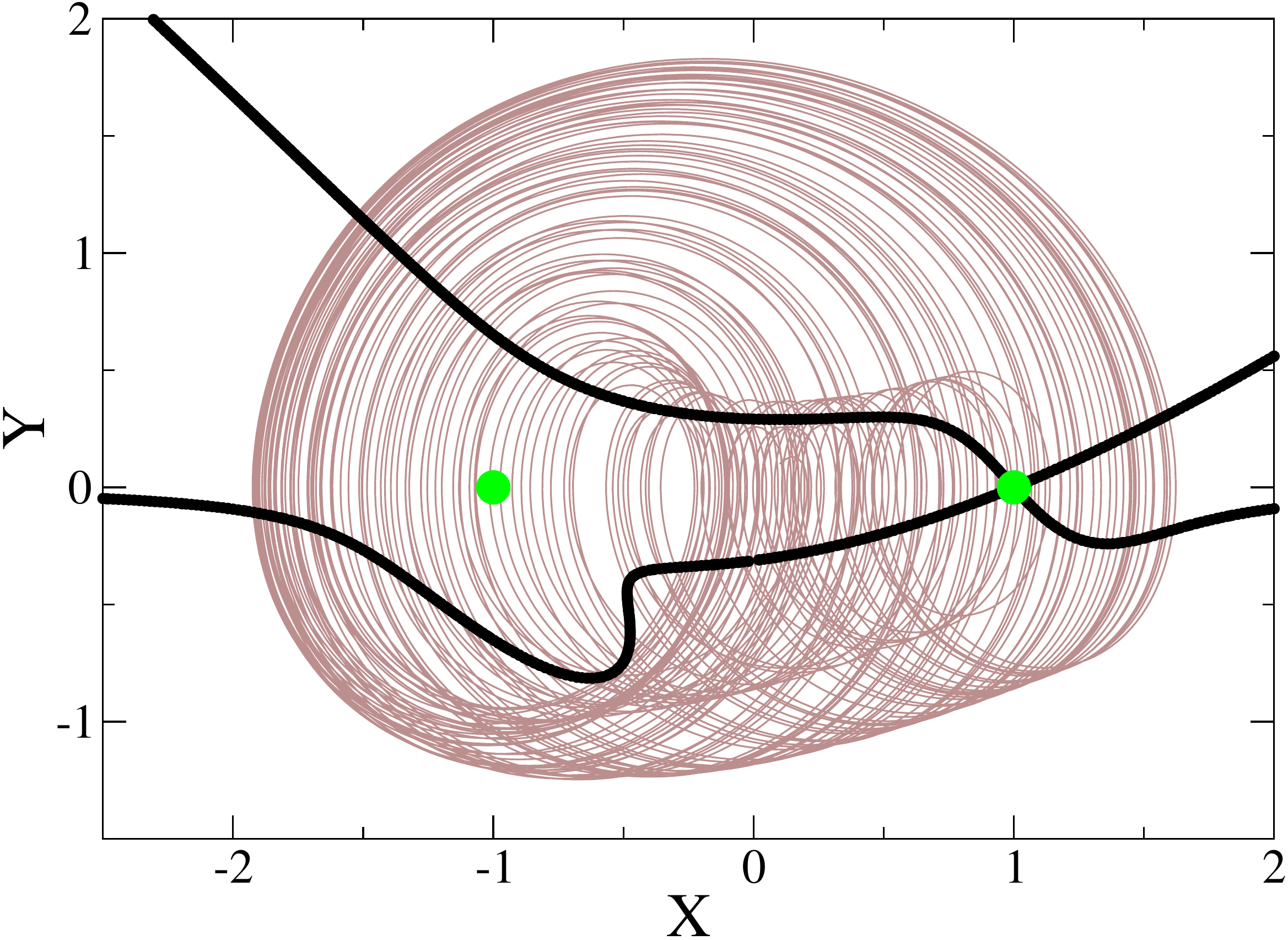}}
\subfigure[$(A,B,C,R):=(-1.3,-2.6,-1.7,1)$]{
  \includegraphics[height=6cm,width=\columnwidth]{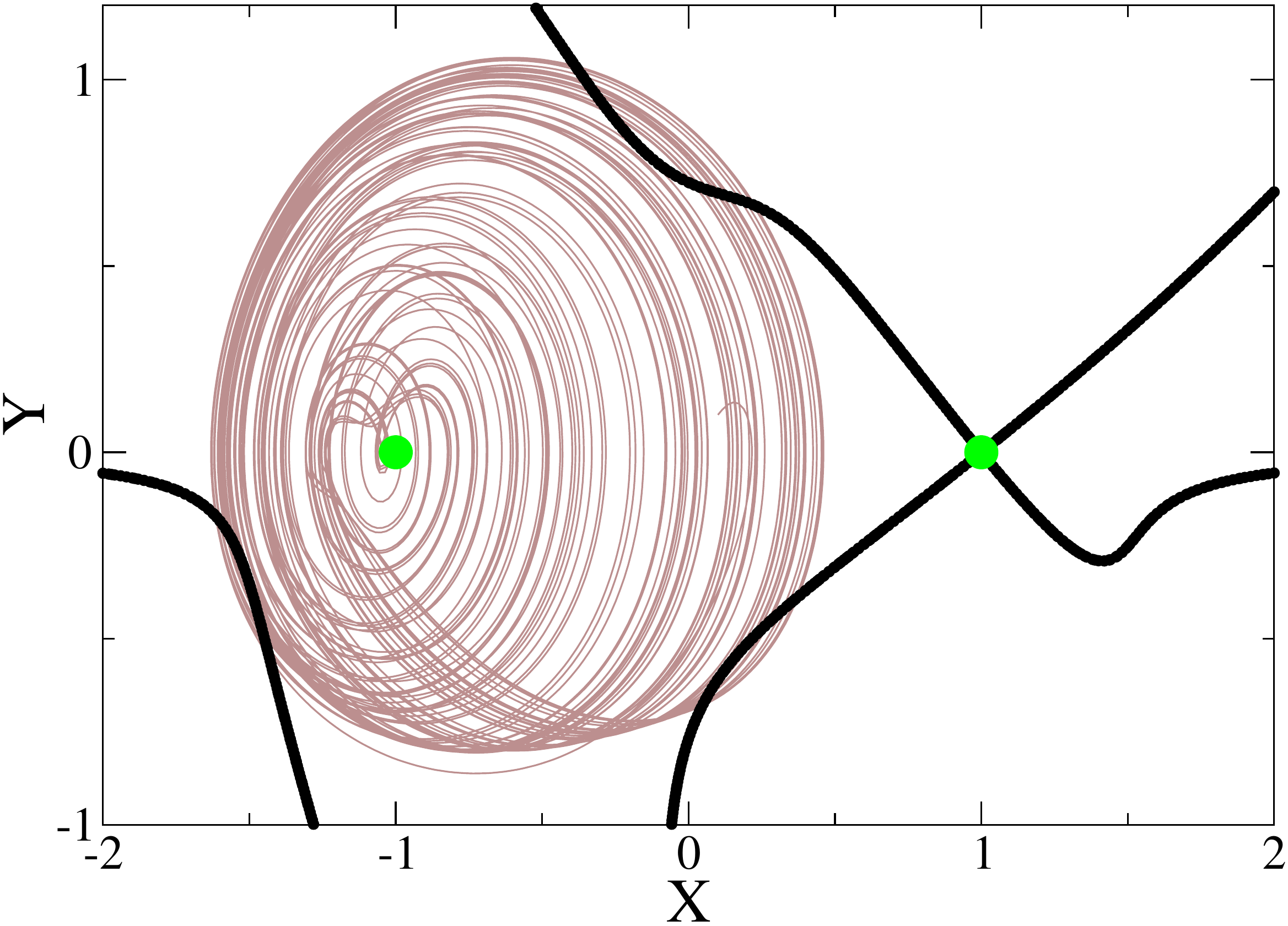}}
\caption{$X-Y$ projection of the connecting curves, fixed points and phase space trajectories from Eq. (\ref{eq:4dode}) with $G_1(x;R)=x^2-R$.  (a) Both fixed points have index $\kappa=1$.  The phase space dynamics near the fixed points are described by a stable limit cycle.  (b-c) Asymmetries in the vortex indices result in a fixed point with $\kappa=2$ (left) and one with $\kappa=1$ (right).  The structure of the strange attractors is strongly influenced by the resulting vortices and hypervortices.}
  \label{fig:4d2fp}
\end{minipage}
\end{figure}

A stable limit cycle is produced for $R=0.15$.  The fixed points are 
located at $x_{f_1}=-\sqrt{R}$ and $x_{f_2}=\sqrt{R}$.  They have coordinates 
$(a,b,c)=(-B,-A,-2x_f)$ in the catastrophe control parameter space 
shown in Fig. \ref{fig:A4}(c).  Both fixed points reside in a 
region where they have a single pair of complex conjugate eigenvalues and 
two real eigenvalues ($\kappa_{1,2}=1$).  The connecting curves and limit 
cycle for this case are shown in Fig. \ref{fig:4d2fp}(a).  The $n-2\kappa=2$ 
vortex core curves are observed to pass through each of the fixed points.  

As $R$ is increased, the two fixed points scatter other along the $c$ axis.  
They retain the same value of $\kappa$ until the upper fixed 
point crosses the bifurcation boundary.  The fixed point transition for $R=1$ 
is shown in Fig. \ref{fig:A4}(c).  This value of $R$ generates the strange attractor 
shown in Fig. \ref{fig:4d2fp}(b).  The fixed 
points are located at $x_{f_1}=-1$ and $x_{f_2}=1$ with $\kappa_1=2$ and 
$\kappa_2=1$.  The different $\kappa$ values plays an important role in 
the structure of strange attractor.  Specifically, an asymmetry is 
created because $n-2\kappa_2=0$ connecting curves pass through $x_{f_1}$ 
and $n-2\kappa_1=2$ connecting curves pass through $x_{f_2}$.  

Swirling flow generated near $x_{f_2}$ is transported along these two core curves 
towards $x_{f_1}$.  The process creates a funnel in the center of the 
strange attractor.  As the flow approaches the neighborhood of $x_{f_1}$, it is 
pushed back towards $x_{f_2}$ and the process is repeated.  The two 
vortex core curves that connect to $x_{f_2}$ pass through the funnel 
of the strange attractor, but diverge to wrap around the hypervortex 
near $x_{f_1}$ since they are unable to attach to it.

%%%%%%%%%%%%%%%%%%%%%%%%%%%%%%%%%%%%%%%%%%%%%%%%%
%%%%%%%%%%%%%%%%%%%%%%%%%%%%%%%%%%%%%%%%%%%%%%%%%

Figure \ref{fig:4d2fp}(c) shows an example of a strange attractor that forms in 
the hypervortex region around $x_{f_1}$ for a different set of control parameters 
$(A,B,C,R)=(-1.3,-2.6,-1.7,1)$.  The vortex core curves in this case have 
less influence on the structure of the strange attractor.

%%%%%%%%%%%%%%%%%%%%%%%%%%%%%%%%%%%%%%%%%%%%%%%%%
%%%%%%%%%%%%%%%%%%%%%%%%%%%%%%%%%%%%%%%%%%%%%%%%%

\subsection{$m=3$}

\begin{figure}[htbp] %%%   Fig. 11
\begin{minipage}{\columnwidth}
\centering
\includegraphics[height=6cm,width=\columnwidth]{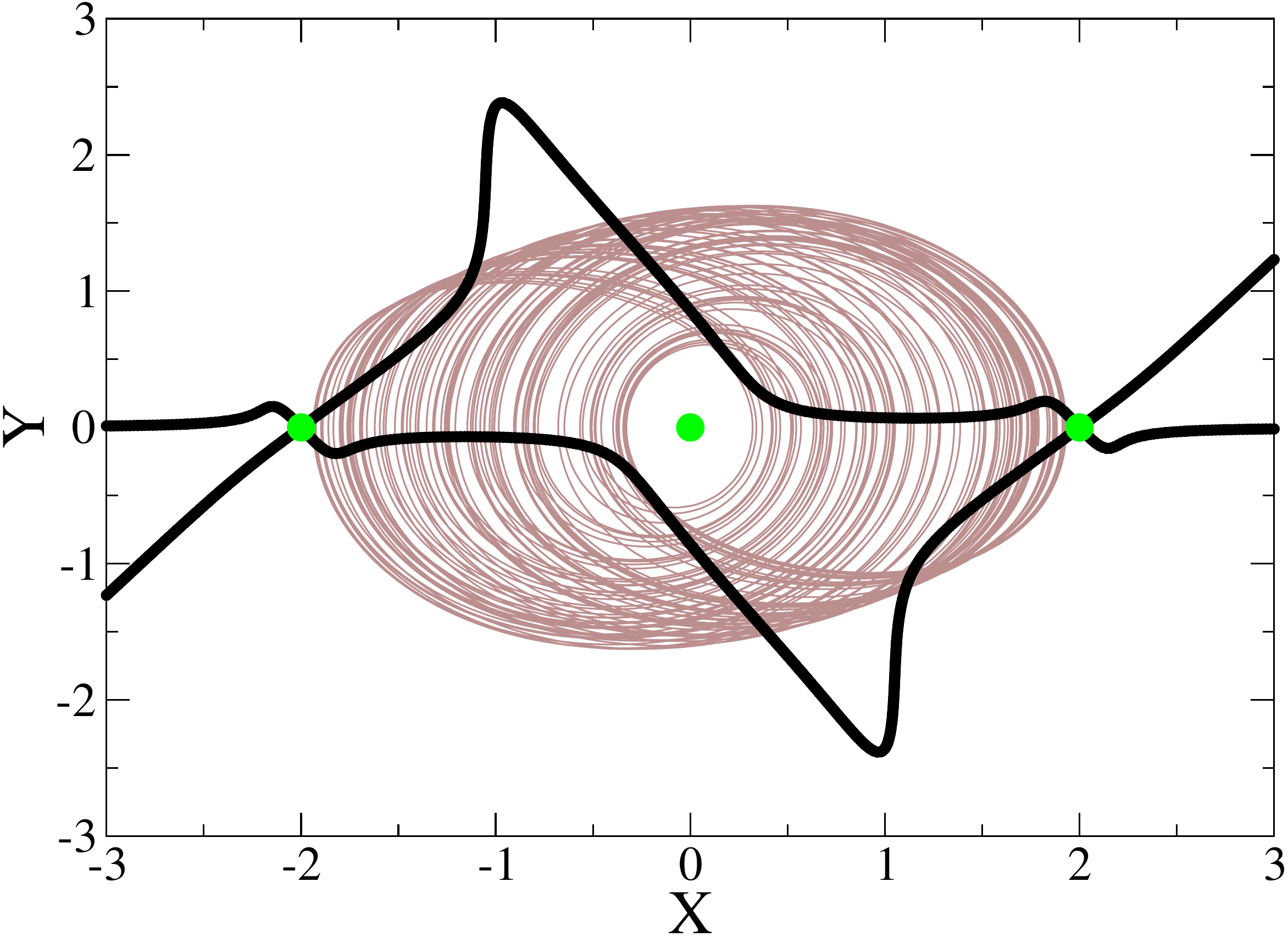}
  \caption{X-Y projection of the strange attractor generated 
    by Eq. (\ref{eq:4dode}) with $G_1(x;R)=x^3-Rx$.  The hypervortex 
    around the origin creates a hole in the attractor and forces the 
    vortex core curves to wrap around it.  Parameter values: $(A,B,C,R)=(-3,-3.2,-1.43,4)$}
  \label{fig:4d3fp} 
\end{minipage}
\end{figure}

For $m=3$ we use $G_1(x;R)=x^3-Rx$ with control parameter 
values $(A,B,C,R)=(-3,-3.2,-1.43,4)$.  The coordinates of the 
fixed points in the catastrophe control parameter space 
are given by $(a,b,c)=(-B,-A,-G_{1}'(x_f;R))$.  Since $a>0$, 
the bifurcation surface splits the control parameter 
space into two regions.  The symmetric fixed points 
share a single coordinate located in the region that 
produces a single pair of complex conjugate eigenvalues ($\kappa=1$) 
and two real eigenvalues.  The fixed point at the origin is 
located in the region that produces two pairs of complex conjugate 
eigenvalues ($\kappa=2$).  It is located in the hypervortex 
near the origin that plays an important role in the structure of the 
strange attractor shown in Fig. \ref{fig:4d3fp}.  The vortex core curves 
pass through the outer fixed points but are unable to enter 
the hypervortex that produces a hole in the center of the attractor.

%%%%%%%%%%%%%%%%%%%%%%%%%%%%%%%%%%%%%%%%%%%%%%%%%
%%%%%%%%%%%%%%%%%%%%%%%%%%%%%%%%%%%%%%%%%%%%%%%%%

\section{Five Dimensions}
\label{sec:5D}

In this section we set $G_2(y,z,u,w;A,B,C,D)=Ay+Bz+Cu+Dw^{3}$ in order to 
study five-dimensional differential dynamical systems that take the 
form 

\begin{equation}
  \begin{array}{l}
    \dot{x} = y \\
    \dot{y} = z \\
    \dot{z} = u \\
    \dot{u} = w \\
    \dot{w} = G_1(x;R)+Ay+Bz+Cu+Dw^{3};
  \end{array}
\label{eq:5dode}
\end{equation}
The characteristic polynomial is

\begin{equation}
\begin{split}
{\rm det}(J-\lambda I_5))=A_5(\lambda)=\\
\lambda^5-C\lambda^3-B
\lambda^2 - A\lambda - G_{1}'(x_f;R) 
\end{split}
\label{eq:cpn5}
\end{equation}
where the substitution $(a,b,c,d)=(-C,-B,-A,-G_1'(x_{f};R))$ 
creates the canonical unfolding of the next catastrophe
in the series of cuspoids $A_5$.  Here again the control 
parameter space is divided into three disjoint open regions 
that describe fixed points with zero, one, or two pairs of 
complex conjugate eigenvalues and a complementary number of 
real eigenvalues.  The open regions are connected and simply 
connected and can be studied as in the case of $A_4(\lambda)$.  
We use $G_1(x_1;R)=x^2-R$ in this section.  

\begin{figure}[htbp] %%%   Fig. 12
\begin{minipage}{\columnwidth}
\centering
\includegraphics[height=6cm,width=\columnwidth]{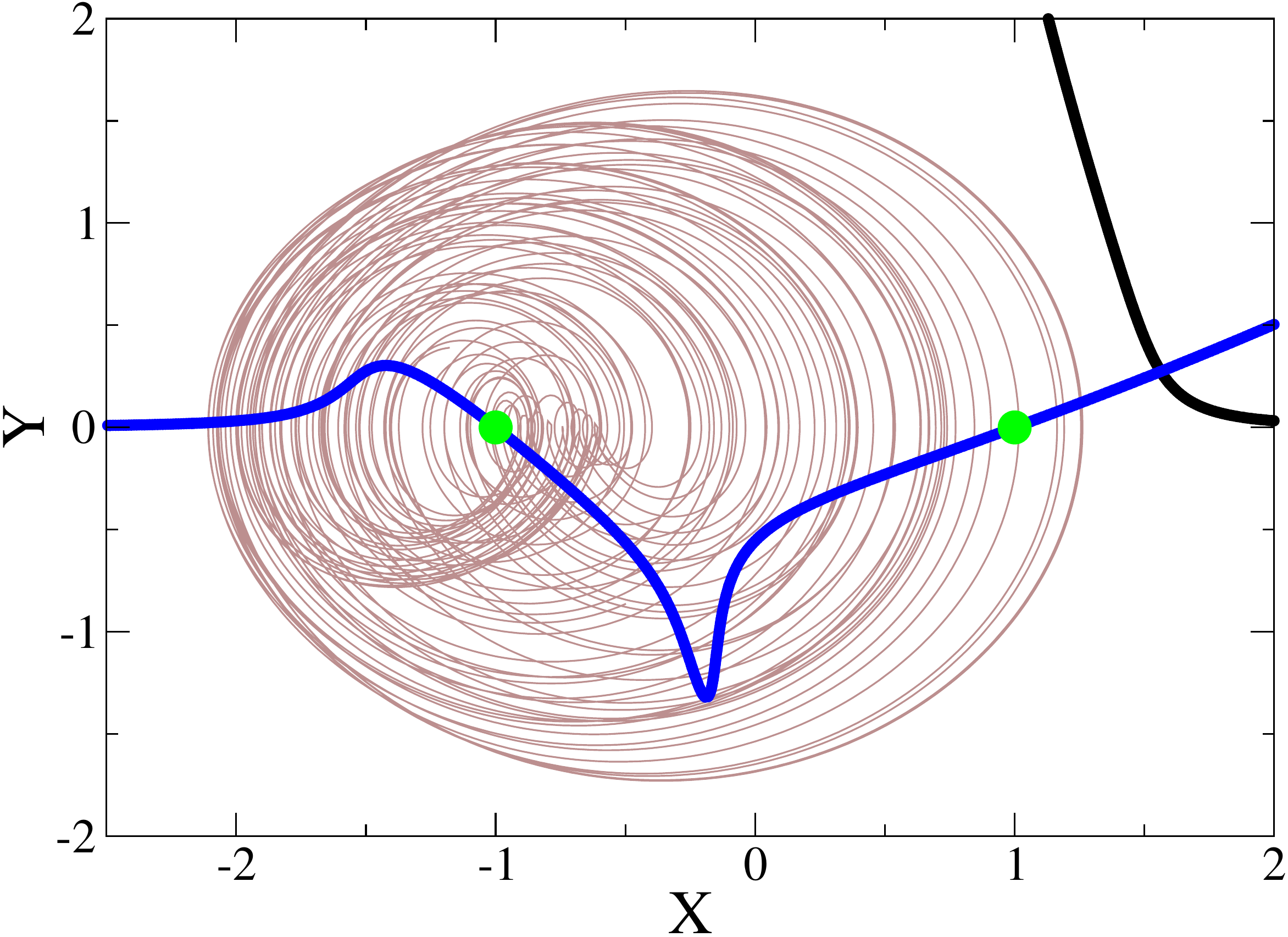}
  \caption{X-Y projection of the strange attractor generated by Eq. 
    (\ref{eq:3dode}) using $G_1(x_1;R)=x^2-R$.  A hypervortex core curve 
    ($\kappa=2$) connects the two fixed points.  
    Parameter values: $(A,B,C,D,R)=(-1.8,-3.8,-3.1,-1,1)$.}
  \label{fig:5d2fp2-2} 
\end{minipage}
\end{figure}

Parameters $(A,B,C,D,R)=(-1.8,-3.8,-3.1,-1,1)$ generate the 
strange attractor shown in Fig. \ref{fig:5d2fp2-2}.  Both 
fixed points have two pairs of complex conjugate eigenvalues 
($\kappa=2$) and $n-2\kappa=1$ hypervortex core curves 
connected to them.  A vortex core curve ($\kappa=1$) is 
also observed to form away from the attractor.  

\begin{figure}[htbp] %%%   Fig. 13
\begin{minipage}{\columnwidth}
\centering
\includegraphics[height=6cm,width=\columnwidth]{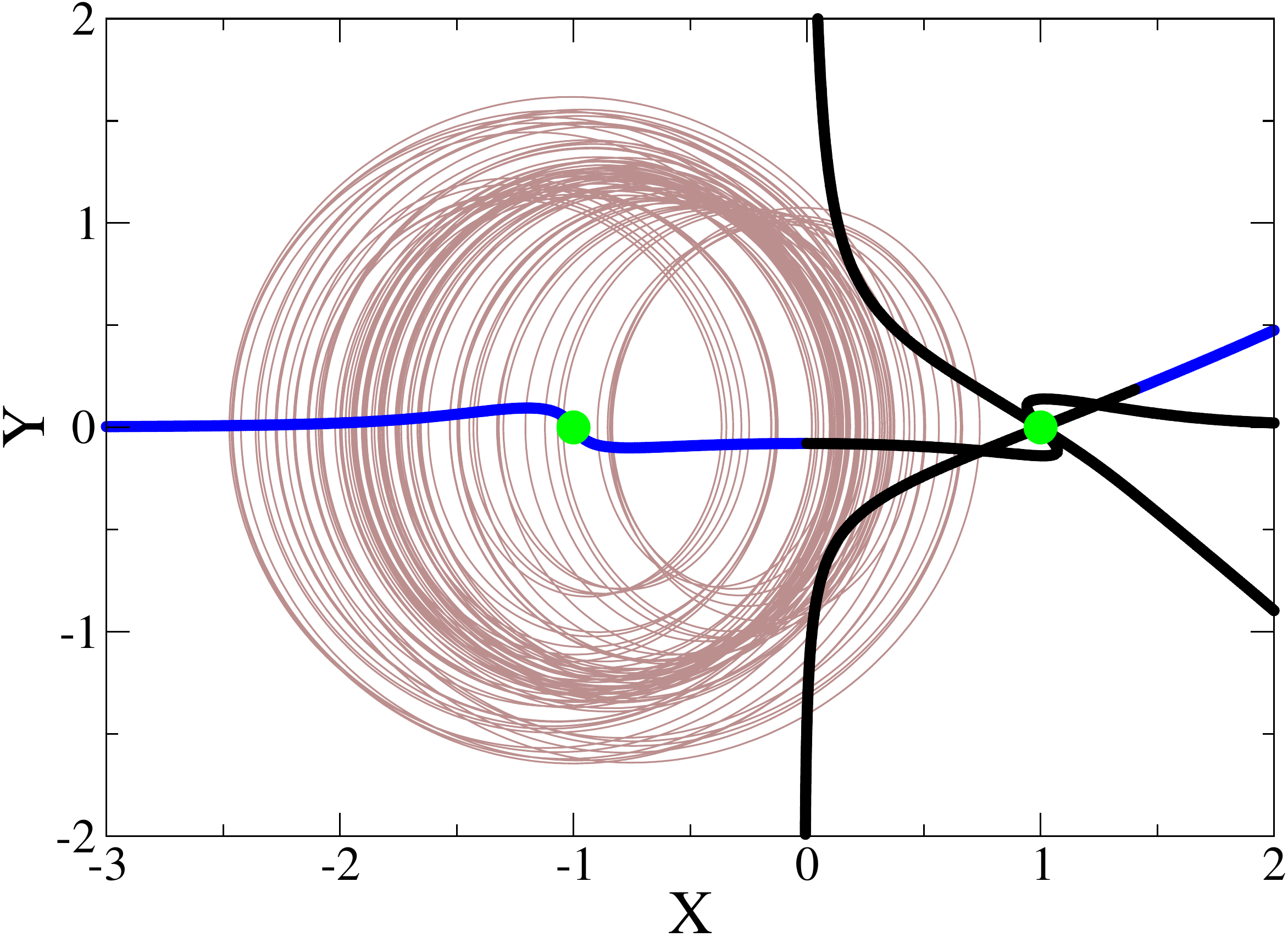}
  \caption{X-Y projection of the attractor generated by Eq. 
    (\ref{eq:3dode}) using $G_1(x_1;R)=x^2-R$.  A single hypervortex core curve ($\kappa=2$) runs 
    through the fixed point on the left.  After a change in stability, three vortex core curves 
    run through the fixed point on the right with $\kappa=1$.  Parameter values: $(A,B,C,D,R)=(-0.7,-7.0,-2.5,-1.0,1.0)$}
  \label{fig:5d2fp2-1}  
\end{minipage}
\end{figure}

Parameters $(A,B,C,D,R)=(-0.7,-7.0,-2.5,-1,1)$ generate the 
strange attractor shown in Fig. \ref{fig:5d2fp2-1}.  The 
fixed point indices $\kappa$ of the two fixed points differ.  
A single hypervortex core curve runs through the fixed 
point $x_{f_{1}}$ with ($\kappa=2$) while $n-2\kappa=3$ vortex 
core curves run through the fixed point $x_{f_{2}}$ with 
($\kappa=1$).

%%%%%%%%%%%%%%%%%%%%%%%%%%%%%%%%%%%%%%%%%%%%%%%%%
%%%%%%%%%%%%%%%%%%%%%%%%%%%%%%%%%%%%%%%%%%%%%%%%%

%%%%%%%%%%%%%%%%%%%%%%%%%%%%%%%%%%%%%%%%%%%%%%%%%
%%%%%%%%%%%%%%%%%%%%%%%%%%%%%%%%%%%%%%%%%%%%%%%%%

\section{Conclusions}
\label{sec:conclusions}

The present work was motivated by an attempt to determine
how invariant sets of greater dimension than 
fixed points can be used to determine
the structure of flows that result from integrating sets of
nonlinear ordinary different equations.  These sets, 
which have been called core curves,
obey the eigenvalue-like equation $J{\bf f}=\lambda {\bf f}$ 
and have previously been used in studies of
three-dimensional flows.  In extending these results to
higher dimensions, we have focused on a special class of
dynamical systems --- differential dynamical systems 
as defined in Eq. (\ref{eq:differentialform}),
which have the same form in all dimensions $n \ge 3$.
Only the driving function $F(x_1, \dots, x_n;c)$ varies
from system to system.

Since all fixed points occur on the $x_1$ axis, it was
convenient to write the driving function as a sum of two
functions, $G_1(x_1;c_1)$ depending only on the coordinate
$x_1$ and a set of control parameters $c_1$, and another
function $G_2(x_2, \dots ,x_n;c_2)$ depending on complementary
variables.  The locations of the fixed points can be put into
canonical form by expressing $G_1$ in terms of a cuspoid
catastrophe $A_m'(x_1;c_1) = x^m + ax^{m-2}+\dots$, where
$m$ is the maximum number of fixed points that occurs under
control parameter variation.  The stability at a fixed point
is determined by the first derivatives $\partial G_2/\partial x_i$,
evaluated at $x_i=0$, $i=2, \dots, n$, and 
$\partial G_1/\partial x_1$.  That is, the 
stability is governed by the linear terms in
$G_2(x_2, \dots ,x_n;c_2)$ and the slope of $G_1$. 
Variations in the stability
of the fixed points is most conveniently studied by
identifying the linearization of $G_2$ with another
cuspoid catastrophe
$A_n'(x_1;c_1) = x^n + ax^{n-2}+\dots $, where $n$ is the
dimension of the dynamical system.  There is a weak
coupling between these two catastrophes given by the
term $G_1'(x_1;c_1)$, which appears as one of the
unfolding parameters for the function $G_2$.  The
slope alternates along the $x_1$ axis from fixed point to
fixed point.

\begin{table}
\caption{Summary of the equations studied.
\label{tab:summary}}
\[  \begin{array}{cccccc}
{\rm Eq.}& n & \# {\rm F.P.} & G_1(x) & G_2 & {\rm Fig.}  \\  \hline
\ref{eq:3dode} & 3 & 2 & x^2-R & Ay+By^2z & \ref{fig:SaddleNode}, \ref{fig:reconnect},\ref{fig:VortStrain} \\
\ref{eq:3dode} & 3 & 3 & x^3-Rx & Ay+By^2z & \ref{fig:3d3fpXY}\\
\ref{eq:4dode} & 4 & 1 & Rx & Ay+Bz+Cu^{3} & \ref{fig:4d1fp}\\
\ref{eq:4dode} & 4 & 2 & x^2-R & Ay+Bz+Cu^{3} & \ref{fig:4d2fp}\\
\ref{eq:4dode} & 4 & 3 & x^3-Rx & Ay+Bz+Cu^{3} & \ref{fig:4d3fp}\\
\ref{eq:5dode} & 5 & 2 & x^2-R & Ay+Bz+Cu+Dw^{3} & \ref{fig:5d2fp2-2},\ref{fig:5d2fp2-1}\\
\end{array}
\]
\end{table}

At a fixed point the core curves are tangent to 
the eigenvectors of $J$ with real eigenvalues.
There are $n - 2\kappa$ real eigenvalues, where
$\kappa$ is the number of complex conjugate pairs of
eigenvalues.  If $n-2\kappa \ge 1$ and $\kappa \ge 1$
then a core curve emanating from the fixed point
may pass through the ``center'' of a
strange attractor and act to organize the flow around it.
If $\kappa = 1$ the curve is called a vortex core curve and
if $\kappa > 1$ it is called a hypervortex core curve.
If $\kappa = 0$ then only strain curves ($n$ of them)
originate at the fixed point.  If $n-2\kappa=0$ then the
fixed point is disconnected from the network of curves
satisfying the defining eigenvalue-like equation.

The eigenvalues of the jacobian at a fixed point
vary as the control parameters vary.
If a real (complex conjugate) pair becomes degenerate
and scatters to become a complex conjugate (real) pair,
then two distinct core curves gradually approach degeneracy
and then detach from (attach to) the fixed point.

The eigenvalues of the jacobian also vary along core
curves.  Changes in stability and/or the number of
real and complex conjugate pairs are closely involved
in rearrangements or recombinations of the core curves.

For differentiable dynamical systems the eigenvalue equation
defining the core curve can be projected to a pair of
constraints acting in a three-dimensional space.
This pair of constraints is given in Eq. (\ref{eq:collapsed1}).
Core curves are not heteroclinic connections in the
dynamical system.  Rather, they satisfy a closely related
set of nonlinear equations that are effectively nonautonomous.
This dynamical system is given in Eq. (\ref{eq:coreequation}).

These ideas have been illustrated for dynamical systems 
with $n=3,m=2,3$, $n=4,m=1,2,3$ and $n=5,m=2$.  
Table \ref{tab:summary} summarizes the cases presented, 
the dynamical system equations studied, and the figures
that illustrate the results.  Supplementary material 
from this work can be found online \cite{supplementary}.  

\section{Acknowledgements}  

We thank Fernando Mut for useful discussion.  

\section{Appendix 1}

Core curves for differential dynamical systems obey the condition
$x_{i+1} = \lambda x_i$, so that $x_k = \lambda^{k-2}x_2$ for
$k=3, \dots, n$.  As a result there are three independent variables
$\lambda, x_1, x_2$ and two constraints:

\begin{equation}
\left[  \begin{array}{cc}
0 & 1 \\
\sum_{j=1}^{n-1} \lambda^{j-1}F_j & F_n \end{array} \right]
\left[  \begin{array}{c}
x_2 \\ F \end{array} \right] =
\lambda \left[  \begin{array}{c}
\lambda^{n-2}x_2 \\
F \end{array} \right]
\label{eq:collapsed1}
\end{equation}
The functions that appear in this collapsed constraint 
equation are obtained by replacing $x_k \rightarrow \lambda^{k-2}x_2$.
The general form for the vortex core curve is independent
of the dimension $n$ of the differential dynamical system
and is described by a curve in the space $R^3$ with
coordinates $(\lambda,x_1,x_2)$.

A dynamical equation that core curves satisfy has the form

\begin{equation}
\left( \frac{\partial^2 f_i}{\partial x_j \partial x_k}f_j+
\frac{\partial f_i}{\partial x_j}\frac{\partial f_j}{\partial x_k}
-\lambda \frac{\partial f_i}{\partial x_k}\right) 
\frac{d x_k}{d \lambda} = f_i
\label{eq:coreequation}
\end{equation}

\end{document}